\documentclass[apj]{emulateapj}

\usepackage{hyperref}
\usepackage{CJKutf8}

\begin{document}
\title{SEARCHING FOR BALMER-DOMINATED TYPE Ia SUPERNOVA REMNANTS IN M33}

\author{Chris Ding-Jyun Lin \begin{CJK}{UTF8}{bsmi}(林鼎鈞)\end{CJK}}
\affiliation{Institute of Astronomy and Astrophysics, Academia Sinica, No.1, Sec. 4, Roosevelt Rd., Taipei 10617, Taiwan, R.O.C.\ 
\\ djlin@asiaa.sinica.edu.tw}
\affiliation{Department of Physics, National Taiwan University, Taipei 10617, Taiwan, R.O.C.}

\author{You-Hua Chu \begin{CJK}{UTF8}{bsmi}(朱有花)\end{CJK}}
\affiliation{Institute of Astronomy and Astrophysics, Academia Sinica, No.1, Sec. 4, Roosevelt Rd., Taipei 10617, Taiwan, R.O.C.\ 
\\ yhchu@asiaa.sinica.edu.tw}
\affiliation{Department of Physics, National Taiwan University, Taipei 10617, Taiwan, R.O.C.}

\author{Po-Sheng Ou \begin{CJK}{UTF8}{bsmi}(歐柏昇)\end{CJK}}
\affiliation{Institute of Astronomy and Astrophysics, Academia Sinica, No.1, Sec. 4, Roosevelt Rd., Taipei 10617, Taiwan, R.O.C.\ 
\\ psou@asiaa.sinica.edu.tw}
\affiliation{Department of Physics, National Taiwan University, Taipei 10617, Taiwan, R.O.C.}

\author{Chuan-Jui Li \begin{CJK}{UTF8}{bsmi}(李傳睿)\end{CJK}}
\affiliation{Institute of Astronomy and Astrophysics, Academia Sinica, No.1, Sec. 4, Roosevelt Rd., Taipei 10617, Taiwan, R.O.C.\ 
\\ cjli@asiaa.sinica.edu.tw}


\begin{abstract}
We have searched for Balmer-dominated Type Ia SNRs in M33 by selecting thermal X-ray sources 
with $L_{\rm X} \ge 5\times10^{35}$ ergs s$^{-1}$, identifying associated H$\alpha$ emission 
features, and checking their [\ion{S}{2}] and [\ion{O}{3}] emission properties.  Our search did not 
find any Balmer-dominated Type Ia SNRs in M33. This result is puzzling because M33 is 2--3 times 
more massive than the LMC, yet the LMC hosts five Balmer-dominated Type Ia SNRs and M33 has none.  
We have considered observational biases, interstellar densities and ionization conditions, Type Ia
SN rate expected from the star formation history and Type Ia SN delay time distribution function, and 
metallicity effect.  None of these can explain the absence of X-ray-bright Balmer-dominated Type Ia
SNRs in M33.  It is intriguing that the Galaxy has X-ray-bright and thermal Type Ia SNRs (Kepler and
Tycho) as well as X-ray-faint and nonthermal Type Ia SNRs (G1.9+0.3, SN1006, and RCW86), while
the LMC does not have the X-ray-faint and nonthermal ones and M33 does not have the X-ray-bright
and thermal ones. 
\end{abstract}
\subjectheadings{
galaxies: individual (M33) --- galaxies: ISM --- ISM: supernova remnants
}

\section{Introduction}

Type Ia supernovae (SNe) are recognized to originate from white dwarfs (WDs) in 
binary systems \citep{Whelan1973,Nomoto1982,Webbink1984}; however, it is not 
known whether the binary companion is another WD or a normal star.  If a surviving 
companion can be identified for a Type Ia SN, the nature of its progenitor can be 
affirmed.  

Surviving companions have been searched for in young Type Ia supernova 
remnants (SNRs) because they have not moved far from the sites of SN 
explosion and would be easier to find.  
A handful of young Type Ia SNRs are 
known in the Galaxy (such as Tycho, Kepler, G1.9+0.3, SN1006, and RCW86) and the Large 
Magellanic Cloud (LMC; 0509$-$67.5, 0519$-$69.0, 0548$-$70.4, DEM L71, and N103B), 
but no surviving companions of their SN progenitors have been unambiguously 
identified and confirmed \citep[e.g.,][]{Ruiz-lapuente2004,Ruiz-lapuente2018,
GH2009,GH2012,schaefer2012,edwards2012,kerzendorf2013,
kerzendorf2014,pagnotta2015,Li2017,Li2019,litke2017,kerzendorf2018}.

To enlarge the sample of young Type Ia SNRs, we intend to search for them in the spiral
galaxy M33 at a distance of 820 kpc \citep{Freedman2001}, or a distance modulus (DM) of 24.6.  
For such a distance, the currently being constructed 30m-class optical telescopes will be able to 
detect surviving companions of SN progenitors in the future.

\begin{figure}[h]
    \centering
    \includegraphics[width=0.45\textwidth]{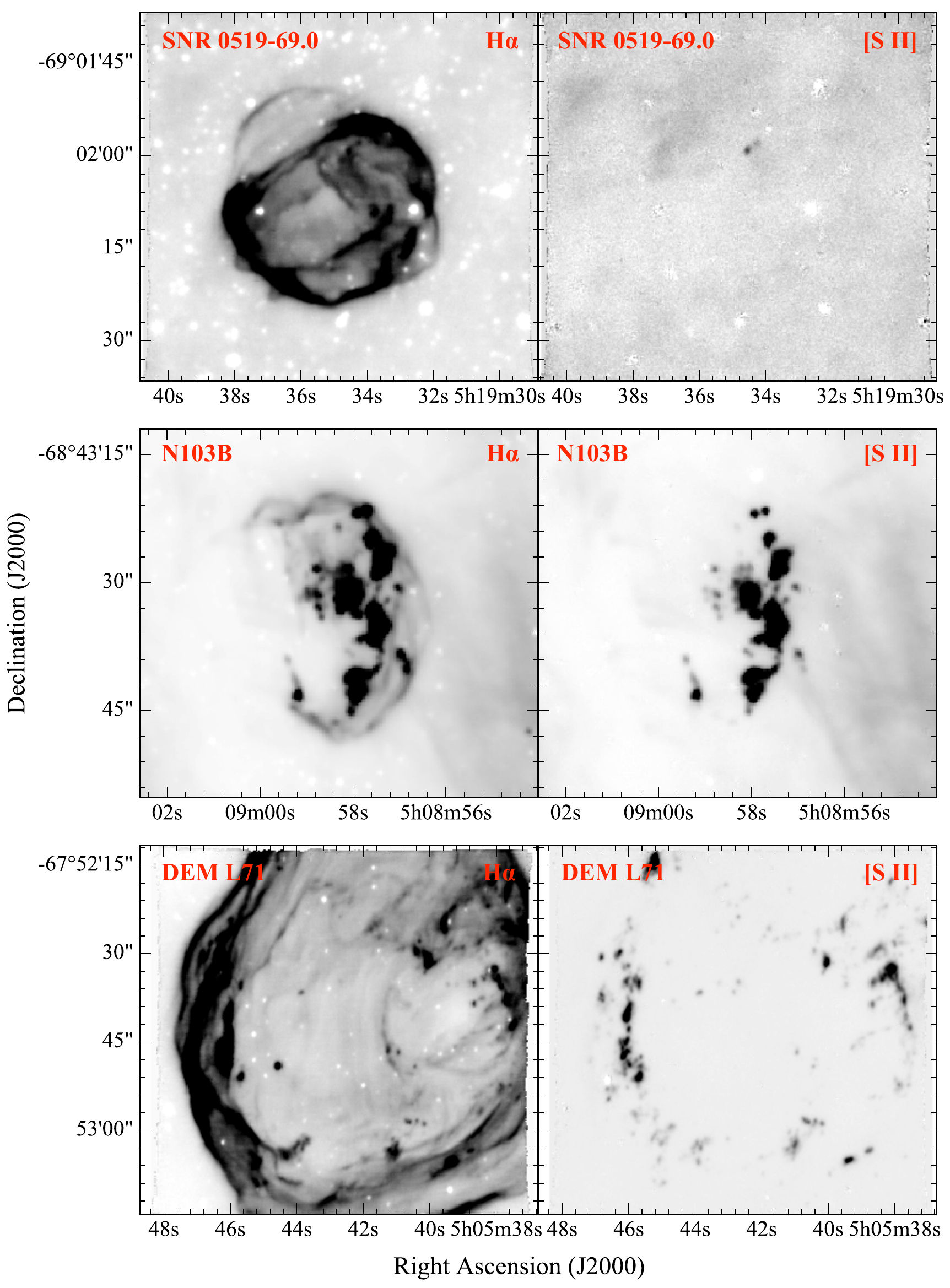}
    \caption{H$\alpha$ and [\ion{S}{2}] images of Balmer-dominated SNRs 0519$-$69.0, N103B, 
    and DEM\L71 extracted from VLT MUSE observations (Li et al. 2020, in preparation). 
    The SNR names are marked in the top left corner of the left panels and the filters are marked 
    in the upper right corner of each panel.  Note that the Balmer-dominated shells disappear in 
    the [\ion{S}{2}] images completely.  Only dense knots are visible in the [\ion{S}{2}] images.}
    \label{fig:LMC3SNRs}
\end{figure}

Young Type Ia SNRs can be identified by their Balmer-dominated optical spectra produced by 
collisionless shocks moving into a partially neutral ambient medium \citep{Chevalier1980}.
{Such Balmer-dominated spectra are not associated with core-collapse (CC) SNRs because their 
surrounding interstellar gas has been photoionized by either their massive progenitors or nearby 
massive stars.  For Balmer-dominated Type Ia SNRs, the ionization fraction of their ambient 
medium is lower and can be used to constrain properties of their SN progenitors \citep{Woods2018}.
The Type Ia nature of Balmer-dominated SNRs has been definitively confirmed through
spectroscopic observations of the light echos of their SNe \citep[e.g.,][]{Krause2008,Rest2008},
or X-ray spectral analyses of their SN ejecta abundances that are consistent with the 
nucleosynthesis yields of Type Ia SNe \citep[e.g.,][]{Hughes1995,Badenes2006,Badenes2008}.}
 
{ Balmer-dominated young Type Ia SNRs are characterized by a Balmer shell, of which the
[\ion{S}{2}] emission is not detected, with [\ion{S}{2}]/H$\alpha$ ratios well below 0.05, and
the [\ion{O}{3}] emission may be detected with [\ion{O}{3}]/H$\beta$ ratios of 0.005--0.015;
however, these SNRs may contain dense knots where the electron densities can be over 
1000 cm$^{-3}$ and [\ion{S}{2}]/H$\alpha$ ratios may range from 0.1 to 0.7
(Li et al.\ 2020, in preparation).   The Balmer shells and dense knots are illustrated
in three example LMC SNRs in Figure~\ref{fig:LMC3SNRs}.

The identification of Balmer-dominated SNRs in galaxies beyond the LMC becomes 
tricky because the Balmer shell and the imbedded dense knots may not be resolved.
Conventional optical surveys of extragalactic SNRs using the selection criterion of 
integrated [\ion{S}{2}]/H$\alpha$ ratio greater than a threshold of 0.4 or higher 
\citep[e.g.,][]{MF1997} would have missed Balmer-dominated SNRs.}
This is the case for optical SNR surveys in M33 \citep{Gordon1998, Lee2014}.  
Thus, we have to make a new search for young 
Balmer-dominated Type Ia SNRs in M33 employing a different methodology.

Using the X-ray and optical properties of Balmer-dominated Type Ia SNRs in the LMC 
as a template, { as shown in Table~1}, we examined optical counterparts of all thermal 
X-ray sources that have X-ray luminosity greater than $5\times10^{35}$ erg s$^{-1}$ 
in M33, but did not find any Balmer-dominated Type Ia SNRs.  This result is puzzling 
because M33 is more massive than the LMC and we expect proportionally more 
Type Ia SNRs.  

This paper reports our
search for Balmer-dominated Type Ia SNRs in M33.  The search methodology is described
in Section 2, the results are presented in Section 3, the SNRs are individually analyzed in 
Section 4, and the implications of the results are discussed in Section 5. A brief summary 
and conclusion is given in Section 6.

\section{Methodology}

To search for young Balmer-dominated Type Ia SNRs in M33, we use known objects in the LMC 
as templates (see Table 1).  We first use an X-ray luminosity threshold to select X-ray sources that 
are as luminous as the Balmer-dominated Type Ia SNRs in the LMC. We further require their 
X-ray spectral properties 
to be consistent with those of thermal plasma emission in order to filter out X-ray binaries (XRBs), 
pulsar wind nebulae (PWNe) and active galactic nuclei (AGNs).  For the selected luminous thermal 
X-ray sources, we examine their optical counterparts, search for nebular shells in H$\alpha$ images 
and check [\ion{O}{3}] and [\ion{S}{2}] images for forbidden line emission.  A luminous thermal X-ray 
source coincident with an H$\alpha$ shell without counterparts in forbidden lines would be a promising 
candidate for Balmer-dominated Type Ia SNR.

\subsection{X-ray Luminosity of Young Type Ia SNRs}
 
\begin{deluxetable*}{lccrcc}
\tablecaption{Properties of Balmer-Dominated Type Ia SNRs in the LMC \label{tab:table1}}
\tablehead{
\colhead{SNR} & \colhead{Optical size$^a$} & \colhead{X-ray size$^b$} & \colhead{$L_{\rm X}$ (0.3--2.1keV)$^b$}
   &   \colhead{$kT$$^b$} & \colhead{[\ion{S}{2}]/H$\alpha$$^c$}\\
\colhead{} & \colhead{(pc)} & \colhead{(pc)} & \colhead{(10$^{36}$ ergs s$^{-1}$)} &  \colhead{(keV)} &}
\startdata
0509$-$67.5  & 7.8$\times$7.1 & 8.0$\times$7.5   &  3.4~~~~~~~~~~      &  0.3 & $<$0.02\\
0519$-$69.0  & 9.7$\times$8.3 & 8.4$\times$8.1   &  10.6~~~~~~~~~~    &  0.4, 0.6 & $<$0.03\\
DEM L71       & 23$\times$17   & 21$\times$18     &  7.7~~~~~~~~~~      &  0.2, $\sim$0.8 & $\sim$0.1\\
0548$-$70.4  & 28$\times$25   & 29$\times$26     &  0.8~~~~~~~~~~      &  0.6 & ...\\
N103B           & 7.7$\times$6.2 & 7.5$\times$7.4    &  16.7~~~~~~~~~~  &   $\sim$ 1 &  0.1--0.2
\enddata
\tablecomments{\\$^a$ Measured from \emph{HST} images, adopting
1$''$ = 0.25 pc for the LMC.\\ 
$^b$ X-ray size, luminosity $L_{\rm X}$, and plasma temperature $kT$ are from Chandra Supernova Remnant Catalog 
\href{https://hea-www.harvard.edu/ChandraSNR/snrcat_lmc.html}{(https://hea-www.harvard.edu/ChandraSNR/snrcat{\_}lmc)}.
The plasma temperature is for the bulk soft X-ray emission that includes the Fe L-shell lines.  The plasma temparature
of N103B is from \citet{Lewis2003}.  \\
$^c$ Integrated over the entire SNR.  From Li et al.\ 2020, in preparation.  The background diffuse ISM emission has 
    high [\ion{S}{2}]/H$\alpha$ ratios, and a highly non-uniform background causes large uncertainty in the measurements. }
\end{deluxetable*}


\begin{figure*}
    \centering
    \includegraphics[width=0.8\textwidth]{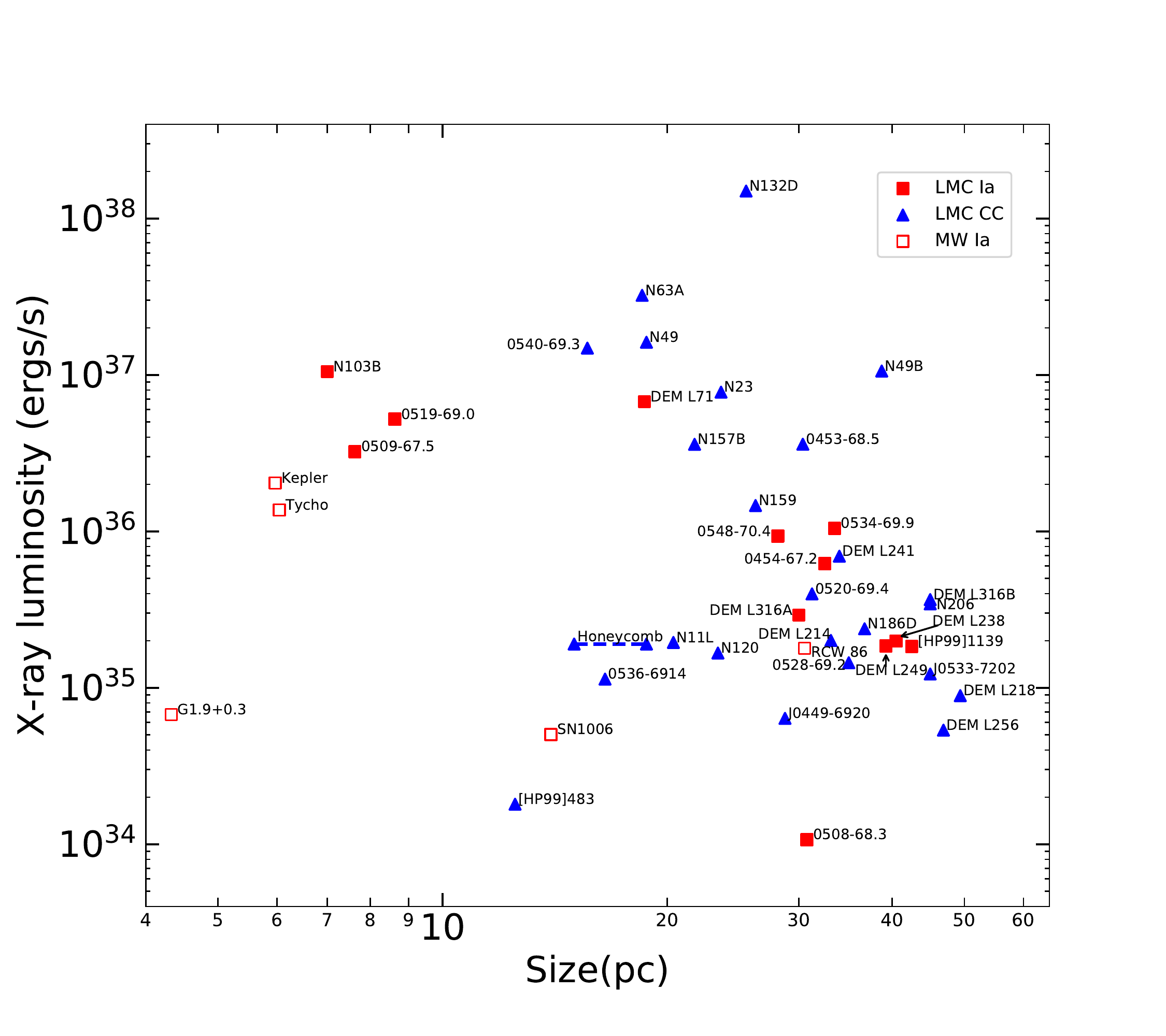}
    \caption{X-ray luminosity versus size plot for LMC SNRs, reproduced from \citet{Ou2018}.  
    Young Type Ia SNRs in the Milky Way galaxy are also included in this plot for comparison.
    The core-collapse (CC) SNRs are plotted in triangles, and the Type Ia SNRs in squares, with
    filled symbols for the LMC and open symbols for the Milky Way.  SNRs greater than 50 pc
    in diameter are not included in this plot.}
    \label{fig:LxSizeLMCSNRs}
\end{figure*}

The LMC hosts five Balmer-dominated Type Ia SNRs.  The first four were
reported by \cite{Tuohy1982}: 0509-67.5, 0519-69.0, DEM L71, and 0548-70.4.  
The latter two are larger and show traces of forbidden line emission, which has been 
interpreted to be caused by an age effect, as the post-shock material gradually becomes 
collisionally ionized. The fifth, N103B, has a Balmer-dominated shell encompassing dense 
ionized knots of circumstellar medium (CSM) that emit bright forbidden lines.  
Examples of Balmer shells and dense knots of three LMC SNRs are shown in 
Figure \ref{fig:LMC3SNRs}.  
The existence of dense CSM in a Type Ia SNR may imply that the WD progenitor had a 
normal star companion \citep{Hachisu2008}.  The optical and X-ray sizes and X-ray 
luminosities ($L_{\rm X}$) in the 0.3--2.1 keV band of these 5 Balmer-dominated Type Ia 
SNRs are given in Table \ref{tab:table1}.  
{ Their unabsorbed $L_{\rm X}$ versus size are plotted in 
Figure~\ref{fig:LxSizeLMCSNRs}, where other LMC SNRs and Galactic 
young Type Ia SNRs are also included for comparisons \citep{Ou2018}. 
The $L_{\rm X}$ of the four smaller Balmer-dominated LMC SNRs are 
all greater than 10$^{36}$ ergs s$^{-1}$, while the largest Balmer-dominated
LMC SNR is fainter, with $L_{\rm X} \sim 8\times10^{35}$ ergs s$^{-1}$.   
The small Galactic Type Ia SNRs appear to show two distinct categories:
the X-ray-bright ones have mostly thermal emission with $L_{\rm X}$ 
greater than 10$^{36}$ ergs s$^{-1}$, such as the Kepler and Tycho SNRs
\citep{Badenes2007}, and the X-ray-faint ones have mostly nonthermal emission
with $L_{\rm X}$ less than $2\times10^{35}$ ergs s$^{-1}$, such as G1.9+0.3, 
SN1006, and RCW86 \citep{Reynolds2008,Koyama1995,Williams2011}.
The X-ray-bright ones are similar to those in the LMC, while the X-ray-faint ones
do not have counterparts in the LMC. 
In this work, we use the criterion of $L_{\rm X} \ge 5\times10^{35}$ ergs s$^{-1}$ to select 
candidates of X-ray-bright Balmer-dominated Type Ia SNRs in M33. }

\begin{figure*}[ht]
    \centering
    \includegraphics[width=0.95\textwidth]{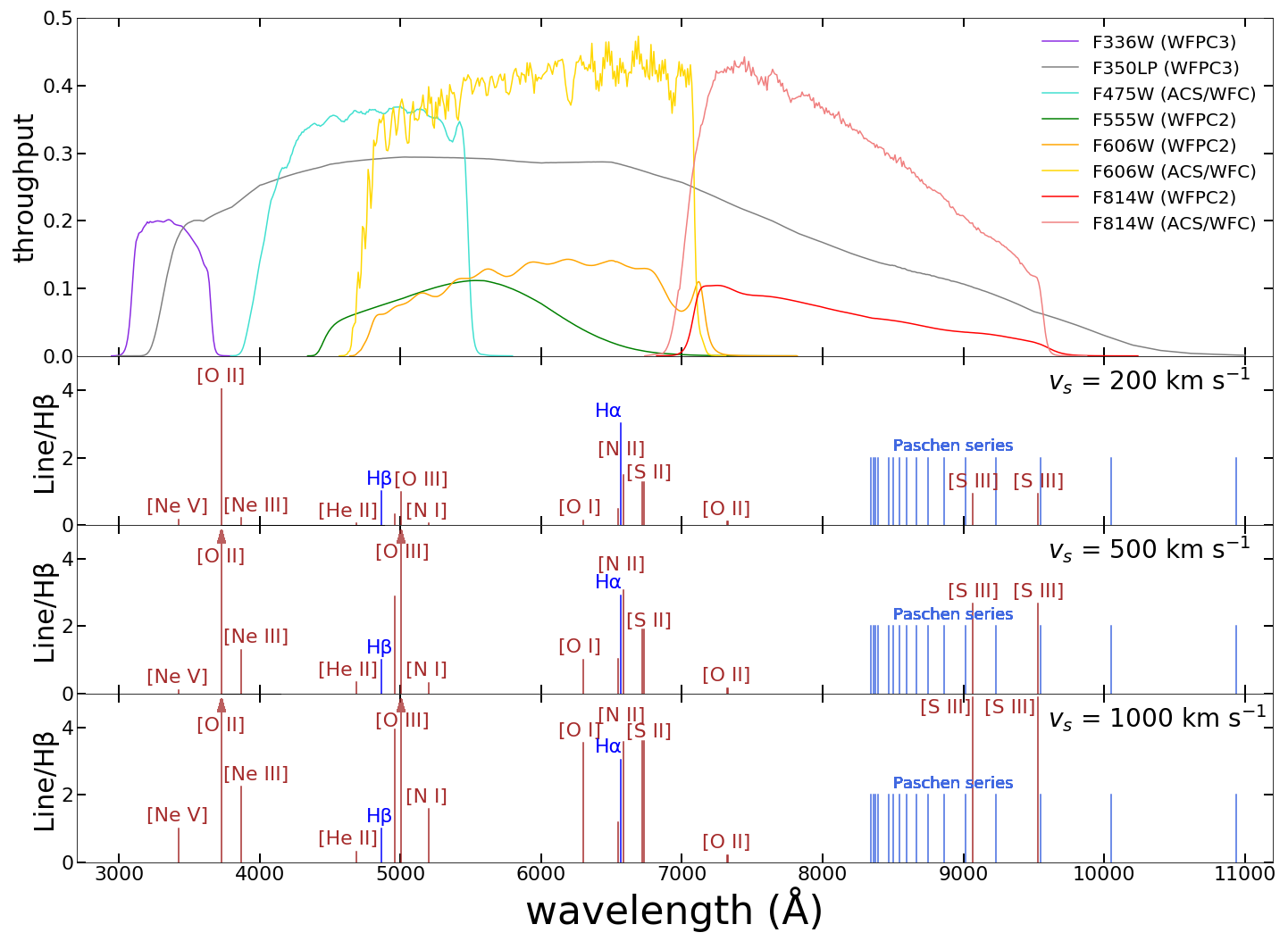}
    \caption{\emph{HST} filter profiles and stronger emission lines for shock components of solar 
    abundance model from \cite{Allen2008} with shock velocity $v_{\rm s}$ = 200, 500, 1000 km s$^{-1}$. 
    The emission lines which have an arrow on their top indicate their line strengths are out of scale.}
    \label{fig:BandsCompare}
\end{figure*}

M33 has been observed by \emph{Chandra X-ray Observatory} for a total of 1.4 Ms 
in the program ChASeM33 \citep{Tullmann2011}.  These ACIS-I observations cover 
about 70\% of the central area of radius $\sim$ 4 kpc, and can detect sources with 
$L_{\rm X} > 4\times10^{34}$ ergs s$^{-1}$ in the 0.5--2.0 keV band.  The brightest 
sources have adequate counts in at least 8 spectral bins to allow meaningful spectral 
fits to determine $L_{\rm X}$, and the best-fit results are reported in Table 9 of 
\citet{Tullmann2011}.  They have considered two spectral models: power-law for 
nonthermal sources and vapec (Variable abundance Astrophysical Plasma Emission 
Code) for thermal plasma sources.  { To search for X-ray-bright Balmer-dominated 
Type Ia SNRs}, we have selected all thermal X-ray sources whose vapec model fits 
indicate unabsorbed $L_{\rm X} \ge 5\times10^{35}$ ergs s$^{-1}$. 
Twenty-one sources meet these selection criteria.

\subsection{Balmer-dominated Shell-like Structure}
Balmer-dominated Type Ia SNRs in the LMC all exhibit a shell-like structure in H$\alpha$ 
images, and the H$\alpha$ shells essentially disappear in [\ion{S}{2}] $\lambda\lambda$6716, 6731, 
and [\ion{O}{3}] $\lambda$5007 images, as illustrated in Figure \ref{fig:LMC3SNRs}.  
This contrast can be explained by collisionless shocks advancing into a partially neutral 
medium, where the [\ion{O}{3}] and [\ion{S}{2}] line emission is expected to be 10-100 times 
fainter than the H$\alpha$ emission \citep{Chevalier1980}.

We have used the H$\alpha$, [\ion{S}{2}], and [\ion{O}{3}] images obtained with the 
Mosaic CCD camera on the Mayall 4-m Telescope at the Kitt Peak National Observatory 
(KPNO) on 2000 October 5 by \citet{Massey2007}.  Their M33-Center field, centered at 
01h33m50.90s, $+$30$^\circ$39$'$37\farcs{0} with a $36'\times36'$ field of view, covers all 
the \emph{Chandra} X-ray sources that have thermal plasma-emission in M33. Each line 
image was taken with a filter of $\sim$50 \AA\ width, and five 300 s exposures dithered to 
compensate CCD gaps. The H$\alpha$ image is used to search for shell structure associated 
with the X-ray source and the [\ion{O}{3}] and [\ion{S}{2}] images are used to check their 
forbidden line emission.

At the distance of M33, 1$''$ corresponds to roughly 4 pc, and ground-based
images with $\sim1''$ seeing are inadequate to resolve the SNR shells.  We have
thus searched for archival \emph{Hubble Space Telescope (HST)} images of M33 to
complement the above-mentioned 4-m Mosaic CCD images.  In the entire M33 galaxy, 
\emph{HST} H$\alpha$ and forbidden line images are available only for the area centered 
on the giant \ion{H}{2} region NGC\,604.  For the other regions in M33, only 
broad-band images are available.  The \emph{HST} broad-band filters often cover 
bright nebular lines as illustrated in Figure \ref{fig:BandsCompare}, where the filter 
response curves are plotted in the top panel and modeled nebular lines for 1000, 500,
and 200 km~s$^{-1}$ shocks \citep{Allen2008} are plotted in the lower three panels, respectively.  
It may be possible to see nebular emission in broad-band images.
We have thus retrieved archival \emph{HST} broad-band images to search for 
nebular emission to examine SNR morphologies at 0\farcs1 resolution.
These will be included in our discussion of individual SNRs.

\section{Nature of the Brightest Thermal X-ray Sources Projected in M33}
In the final source catalog of ChASeM33 \citep{Tullmann2011}, the 21 brightest thermal X-ray sources 
have $L_{\rm X} \ge 5\times10^{35}$ ergs s$^{-1}$. Table \ref{tab:table2} lists their source No., source ID, 
absorption column densities, plasma temperatures, $L_{\rm X}$, other names, identifications of their nature, 
detection in forbidden line images, and optical sizes. Among these 21 sources, nine are associated with 
known SNRs \citep{Long2010, Lee2014}.  These SNRs are individually discussed in Section 4. 
We have also examined the other 12 sources in detail.  As described 
below, we confirm 5 foreground stars (FSs) in the Galaxy, suggest 2 high-mass X-ray binaries (HMXBs) in 
M33, and comment on the other 5 sources.  The plasma temperature distributions of
these different categories of objects are shown in Figure \ref{fig:PTC}.  It is clear from this temperature 
distribution that the unidentified sources are unlikely to be SNRs because their plasma temperatures are too high.

\begin{deluxetable*}{rcccrcccc}
\tablecaption{21 Brightest Thermal X-ray Sources in M33 \label{tab:table2}}
\tablehead{
\colhead{No.$^{a}$} & \colhead{Source ID$^{a}$} & \colhead{$N_{\rm H}\,^{b}$} & \colhead{$kT^{c}$} & \colhead{$L_{\rm X}$(0.35-2 keV)$^{a}$}
 & \colhead{Other Names$^{d}$} & \colhead{Identification$^{e}$} & \colhead{Detection$^{f}$} & \colhead{Optical Size}\\
\colhead{} & \colhead{} & (10$^{22}$ cm$^{-2}$) & \colhead{(keV)} &\colhead{(10$^{35}$ ergs s$^{-1}$)} & \colhead{(L10; LL14)} 
  & \colhead{} & \colhead{} & \colhead{(pc)}}
\startdata
  68 & 013303.55+303903.8 & 0.42 & 3.75 &  8.3~~~~~~~~~    & ...            & HMXB           & [SII], [OIII] & ... \\
100 & 013311.09+303943.7 & 0.01 & 0.55 &  5.2~~~~~~~~~    & 023; 035 & SNR              & [SII], [OIII] & 33$\times$27\\
102 & 013311.75+303841.5 & 0.01 & 0.55 &  110.6~~~~~~~~~ & 025; 037 & SNR             & [SII], [OIII] & 21\\
184 & 013329.04+304216.9 & 0.01 & 0.56 &  19.4~~~~~~~~~  & 036; 061 & SNR             & [SII], [OIII] & 18$\times$11\\
188 & 013329.45+304910.7 & 0.01 & 0.56 &  10.8~~~~~~~~~  & 037; 062 & SNR             & [SII], [OIII] & 34\\
197 & 013331.25+303333.4 & 0.01 & 0.60 &  24.0~~~~~~~~~  & 039; 067 & SNR             & [SII], [OIII] & 28$\times$24\\
236 & 013335.90+303627.4 & 0.01 & 0.77 &  8.0~~~~~~~~~    & 045; 074 & SNR             & [SII], [OIII] & 48$\times$30\\
245 & 013337.08+303253.5 & 1.78 & 0.16 &  112.1~~~~~~~~~ & 046; 076 & SNR             & [SII], [OIII] & 45$\times$36\\
279 & 013341.90+303848.8 & 0.01 & 0.74 &  12.9~~~~~~~~~   & ...            & FS (M4--5\,V)  & [SII], [OIII] & ... \\
287 & 013343.39+304630.6 & 0.01 & 0.48 &  11.9~~~~~~~~~   & ...            & FS (F2--5\,V?)   & [SII], [OIII] & ... \\
301 & 013346.81+305452.8 & 1.51 & 1.05 &  16.5~~~~~~~~~   & ...            & FS (K1--6\,V)   & [SII], [OIII] & ... \\
320 & 013352.12+302706.5 & 0.04 & 4.67 &  8.9~~~~~~~~~    &  ...            & HMXB          & [SII], [OIII] & ... \\
334 & 013354.91+303310.9 & 0.01 & 0.44 &  20.4~~~~~~~~~  & 071; 107  & SNR             & [SII], [OIII] & 21\\
427 & 013410.69+304224.0 & 0.01 & 0.31 &  11.8~~~~~~~~~   & 096; 140 & SNR             & [SII], [OIII] & 22\\
450 & 013416.76+305101.8 & 0.68 & 4.34 &  5.6~~~~~~~~~     & ...            & NOC            & ...               & ... \\
462 & 013418.22+302446.1 & 0.29 & 0.81 &  6.7~~~~~~~~~     & ...            & FS (K6\,V)        & [SII], [OIII] & ... \\
470 & 013421.09+304932.3 & 4.84 & 2.19 &  17.9~~~~~~~~~   & ...            & NOC            & ...              & ... \\
528 & 013432.02+303454.1 & 3.02 & 10.87&  20.7~~~~~~~~~  & ...            & NOC            & ...              & ... \\
585 & 013444.23+304920.3 & 0.48 & 0.72 &  9.1~~~~~~~~~     & ...            & FS (M4\,V)  & [SII], [OIII] & ... \\
587 & 013444.62+305535.0 & 0.97 & 3.58 &  24.5~~~~~~~~~   & ...            & HMXB?        & [SII]          & ... \\
608 & 013451.10+304356.7 & 0.85 & 2.52 &  5.4~~~~~~~~~     & ...            & NOC             & ...             & ... \enddata
\tablecomments{\\$^{a, b, c}$ From ChASeM33 by \cite{Tullmann2011}.\\ $^{b}$ Column density internal to M33.\\ $^{c}$ Plasma temperatures.\\ $^{d}$ L10-nnn denotes SNRs from \citet{Long2010}, L14-nnn denotes SNRs from \citet{Lee2014}.\\ $^{e}$ Physical nature of the source concluded in this paper. SNR -- Supernova Remnant; HMXB -- High Mass X-ray Binary; NOC -- No Optical Counterpart; FS -- Foreground Star with spectral type given in parentheses. \\ $^{f}$ Detection in forbidden line images. Note that stellar continuum of bright stars can also be detected.
}
\end{deluxetable*}

\subsection{Foreground Stars}
Five of these 21 brightest thermal X-ray sources, No.\ 279, 287, 301, 462, and 585, are 
associated with known FSs \citep{Tullmann2011}.  
{ The parallaxes of four of these stars can be found in the Gaia data release 2 
\citep[DR2,][]{Gaia2018} to confirm their being in the solar neighborhood.}
Their spectral types and origins of X-ray emission can be assessed from 
their photometric measurements, assuming negligible extinction and 
comparing them with those of main sequence stars compiled in Table 5 of \citet{Pecaut2013}.

Source 279's stellar counterpart has $V$ = 17.3, $B-V$ = 1.73, $V-R$ = 1.37
\citep{Massey2006, Massey2007, Massey2016}, and $H-K$ = 0.308 \citep{Cutri2003}.
{ Gaia DR2 reports a parallax of 10.789$\pm$0.129 mas, or a distance of 93$\pm$1 pc.
Both the absolute magnitude and colors of this star are consistent with those of a main
sequence M4--M5 star.  Its $L_{\rm X}/L_{\rm bol}$ is $\sim$4$\times$10$^{-4}$, 
consistent with the stellar coronal origin of the X-ray emission \citep{Fleming1995,Zickgraf2005}.}
This X-ray source has also been marked as variable by \citet{Tullmann2011}.

Source 287's stellar counterpart was not reported by \citet{Massey2006, Massey2007, Massey2016}.  
Other catalogs report $V$ = 11.4, $B-V$ = 0.37 to 0.43, $V-R$ = 0.25 to 0.67, $V-I$ = 1.22 
\citep{Zacharias2004,Ivanov2008,Bourg2014}, $J-H$ = 0.34 and $H-K$ = 0.11 \citep{Cutri2003}. 
The large $V-R$ differences between observations made by different people at different epochs 
indicate that this star might be variable or have an eclipsing companion.  { No main sequence stars
can meet the observed colors from blue to near-IR wavelengths.  The $B-V$ and $V-R$ colors
suggest a main sequence spectral type of F2--F5\,V; however, the $V-I$ color suggests K5\,V.
Gaia DR2 reports a parallax of 5.224$\pm$0.067 mas, or a distance of 191$\pm$3 pc.
An F2--F5\,V star at such a distance would have $V$ $\sim$ 10, brighter than the observed 11.4.
We conclude that the stellar counterpart of Source 287 is a foreground star, but it is not a single 
main sequence star.}

Source 301 { has two stars projected within 3$''$.  The star nearest to the X-ray position is also
brighter, thus we take this star as the optical counterpart.}  It has  $V$ = 14.4,
$B-V$ = 1.08, $V-R$ = $-$0.27, and $U-B$ = 1.03
\citep{Massey2006,Massey2007,Massey2016}, or $V$ = 13.89, $B-V$ = 1.25 and $V-R$ = 0.465 
\citep{Qi2015}, and $J-H$ = 0.58, $H-K$ = 0.08 \citep{Cutri2003}. The optical colors and magnitudes 
reported by the two groups are significantly different; for example, the $V-R$ color even changed sign 
between the two measurements. The colors reported by Massey et al.\ correspond to $\sim$K4\,V, 
if the anomalous $V-R$ is discarded.  \citet{Qi2015} only has three passbands and thus the implied 
spectral type has a larger range of uncertainty, K1--K6\,V.  Assuming a K4 main sequence star, its 
$V$ = 13.9--14.4 implies a distance of 310--240 pc and $L_{\rm X}/L_{\rm bol}$ $\sim$\,2.0--1.3$\times10^{-4}$, 
also consistent with a stellar coronal origin of the X-ray emission.
{ Gaia DR2 did not provide a parallax for this star.}

Source 462's stellar counterpart has $V$ = 16.49, $B-V$ = 1.24, $V-R$ = 0.74, and $R-I$ = 0.74,
\citep{Massey2006,Massey2007,Massey2016}.
These colors and magnitudes are consistent with a K6\,V star at a distance of $\sim$500 pc. 
{ Gaia DR2 reports a parallax of 1.183$\pm$0.060 mas, or a distance of 845$\pm$43 pc,
and an effective temperature ($T_{\rm eff}$) of 4372 K for this star.  This $T_{\rm eff}$ is 
consistent with a K6\,V star. 
Using the photometric distance, we obtain $L_{\rm X}/L_{\rm bol}$  $\sim$ 4$\times$10$^{-4}$, 
consistent with a stellar coronal origin of the X-ray emission.  If the Gaia parallax is used,
$L_{\rm X}/L_{\rm bol}$ would be $\sim$\,1$\times$10$^{-3}$, on the high side of the possible range.}

Source 585 { has two stars projected within 2$''$.  Star 1 is closer, within 0\farcs5, and fainter
with  $V$ =18.31, $B-V$ = 1.66, $V-R$ = 1.37, and $R-I$ = 1.19, while Star 2 is off
by 1\farcs6 and has  $V$ =13.81, $B-V$ = 0.82, $U-B$ =0.47, and $V-R$ = 0.42
\citep{Massey2006,Massey2007,Massey2016}.  The colors and magnitudes of Star 1
are suggestive of an M4\,V star at a distance of $\sim$130 pc, and Star 2 a K0\,V at a 
distance of $\sim$420 pc.  If Star 1 is the only star responsible for the X-ray source,
Its $L_{\rm X}/L_{\rm bol}$ would be $\sim$\,4.6$\times10^{-4}$, consistent with a stellar 
coronal origin of the X-ray emission.  Its distance of 130 pc is much closer than the 500 pc
that corresponds to its Gaia DR2 parallax of 1.979$\pm$0.103 mas.  The Gaia DR2 parallax for
Star 2 is similar, 2.199$\pm$0.024 mas.  
Even if Star 2 contributes to the X-ray emission, the combined $L_{\rm X}/L_{\rm bol}$ is still
a few times 10$^{-4}$, within the range of stellar coronal emission \citep{Fleming1995,Zickgraf2005}.}

\subsection{Candidates for High-Mass X-ray Binaries}

Two of the 21 brightest thermal X-ray sources in M33, Sources 68 and 320, might be HMXBs in M33.

Source 68 has $V$ = 19.45, $U-B$ = $-$0.60, $B-V$ = 0.43, $V-R$ = 0.38, and
$R-I$ = 0.60 \citep{Massey2006,Massey2007,Massey2016}.  The colors do not
agree with any types of stars.  At long wavelengths it appears red and at
short wavelengths it appears blue.  Given M33's DM of 24.6, 
Source 68's $M_V$ would be $-5.2$ even without extinction correction.
It is not red enough to be a red supergiant; thus, it must be an early-type 
massive star in M33.  We suggest that this source is most likely an HMXB in M33.  
Its X-ray luminosity is also consistent with an HMXB on the faint end.

Source 320 has $V$ = 19.43, $B-V$ = 0.35, $U-B$ = $-$0.83, $V-R$ = 0.263, and
$V-I$ = 0.79 \citep{Massey2006,Massey2007,Massey2016}.  It has also been reported
to have $V$ = 18.49, $B-V$ = 1.90, and $V-R$ = $-$1.07 \citep{Qi2015}.  
The discrepancy between these two sets of measurements suggests the source is
variable, and the X-ray source is indeed reported to be variable
\citep{Tullmann2011}. Following the same arguments as for Source 68, we suggest
that the optical counterpart of Source 320 corresponds to a massive early-type star,
and that this source is also an HMXB on the faint end.

Note that the temperatures of the thermal plasma models for the above two sources
are also significantly higher than those of stellar coronal emission (Figure \ref{fig:PTC}).  This lends 
further support to the HMXB explanation.

\subsection{X-ray Sources without Optical Counterparts}

\begin{figure}[t]
    \centering
    \includegraphics[width=0.45\textwidth]{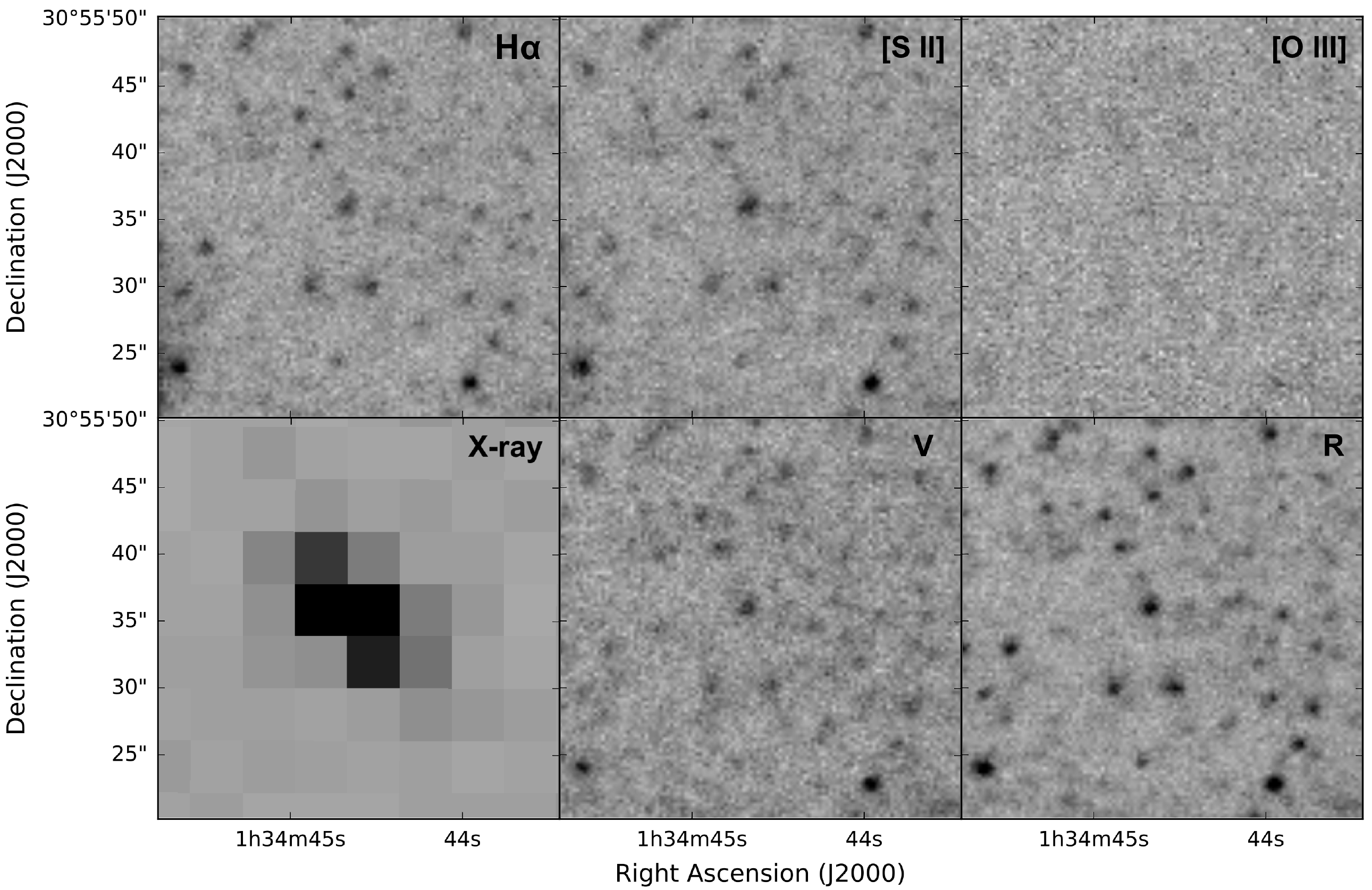}
    \caption{\textbf{No.587, 013444.62+305535.0.} The top panels display
    H$\alpha$, [\ion{S}{2}] $\lambda\lambda$6716, 6731, and [\ion{O}{3}] $\lambda$5007
    images taken with the Mosaic CCD camera on the KPNO 4-m telescope by
    \citet{Massey2007}.  The bottom panels show, from left to right, \emph{Chandra} X-ray 
    image (ObsID 2023, PI: Damiani), and KPNO 4-m Mosaic images in $V$ and $R$.  
    The possible optical counterpart is near the center of the field of view.}
    \label{fig:587FSunid}
\end{figure}

Five of the 21 brightest thermal X-ray sources in M33, Sources 450, 470, 528, 
587 and 608, do not have cataloged optical counterparts.  
We note that Source 587 may have an optical counterpart.
This optical source, located at $\sim$1\farcs2 from the X-ray source, is not 
cataloged by \citet{Massey2006,Massey2007,Massey2016} and must thus be fainter 
than $V$ = 19.5. The X-ray source is located near the periphery of the $Chandra$ observation's field of view, where the point spread function deteriorates rapidly.  
As show in Figure \ref{fig:587FSunid}, a faint point-like source projected near the centroid of the X-ray source is detected in the $V$, $R$, H$\alpha$, and [\ion{S}{2}], but not [\ion{O}{3}], images from \citet{Massey2007}.  
The plasma temperature of this X-ray source is similar to that of Source 68, and
its optical counterpart is somewhat fainter than the optical counterpart of Source 68.
We suggest that Source 587 might also be a HMXB in M33, but this needs to be confirmed
with better X-ray astrometry and optical photometric observations.

\begin{figure}[h]
    \centering
    \includegraphics[width=0.35\textwidth]{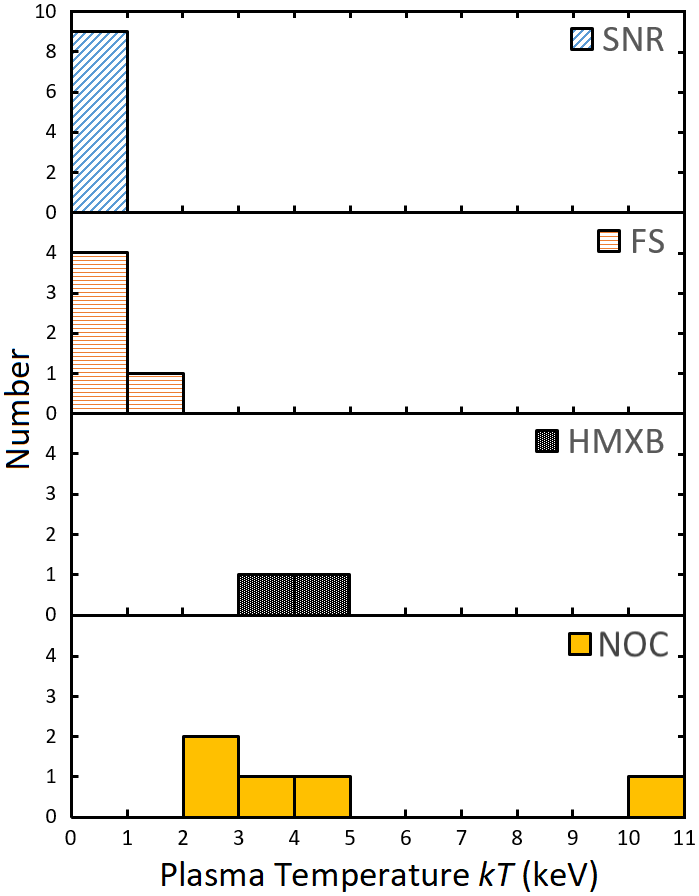}
    \caption{Plasma temperature distributions for the 21 brightest thermal X-ray 
    sources in M33, plotted separately for confirmed SNRs, confirmed FSs,
    candidate HMXBs, and sources with no optical counterparts (NOC).}
    \label{fig:PTC}
\end{figure}

To assess possible nature of the other four bright X-ray sources, we plot a histogram of
the distributions in the model-fit plasma temperature for SNRs, FSs, HMXB candidates, and unidentified objects (including source 587) in Figure \ref{fig:PTC}.
It is clear that all SNRs and foreground stars have low plasma temperatures, with
the highest value at $kT$ = 1.05 keV, while the HMXB candidates have $kT$ around 4 keV.
The five bright X-ray sources without confirmed optical counterparts all have
$kT$ greater than 2 keV.  We suggest that these luminous X-ray sources without
optical counterparts are background AGN or low-mass X-ray 
binaries in M33.  If the optical counterpart of Source 587 can be confirmed, it will 
most likely be a HMXB.

\section{Analyses of Individual Supernova Remnants}

For each SNR, we produce a six-panel figure to show its morphology in different wavelengths.
The top three panels are H$\alpha$, [\ion{S}{2}], and [\ion{O}{3}] images taken with the KPNO 
4-m telescope by \citet{Massey2007}.  The bottom row has an unfiltered
\emph{Chandra} X-ray image (ObsID 6383, PI: Sasaki) on the left, and two \emph{HST} images
taken with various filters from different programs.  Information on the \emph{HST} images are
given in figure captions.

\begin{figure}[h]
    \centering
    \includegraphics[width=0.45\textwidth]{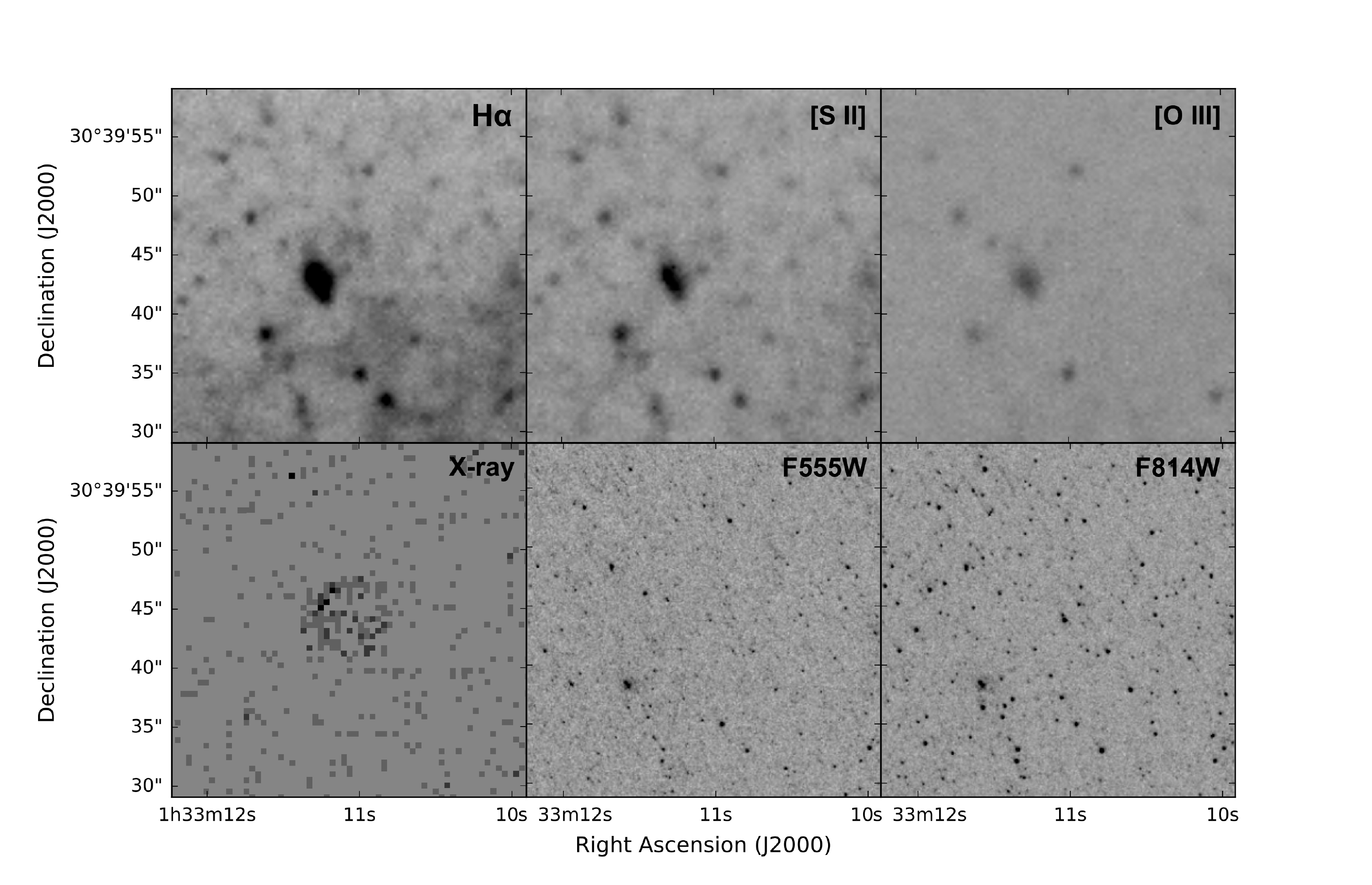}
    \caption{\textbf{No.100, 013311.09+303943.7 (L10-023).}  The \emph{HST} images
    are taken with F555W ($V$) and F814W ($I$) filters in Program 11079 (PI: Bianchi).} 
    \label{fig:100SNR}
\end{figure}

\subsection{No.100, 013311.09+303943.7 (L10-023, LL14-035)(Figure \ref{fig:100SNR})}
The X-ray image of this SNR shows a nice simple shell structure.  The H$\alpha$ image shows
a bright patch of emission along the shell rim in the southeast quadrant and
fainter emission along the northern rim.  The entire SNR is in the northwestern part of the large 
\ion{H}{2} region NGC\,592.  The association with a star-forming 
complex suggests that the SN progenitor is most likely a massive star.  
The large variation of surface brightness along the shell rim is also 
consistent with what is observed in CC SNRs in the LMC 
(Ou et al. 2020, in preparation).
The [\ion{S}{2}] and [\ion{O}{3}] images show a similar morphology of this SNR.
The \emph{HST} broad-band images have too short exposure times to be able to
detect the nebular emission.
We suggest that this is a CC SNR, definitely not a Balmer-dominated Type Ia SNR.

\begin{figure}[h]
    \centering
    \includegraphics[width=0.45\textwidth]{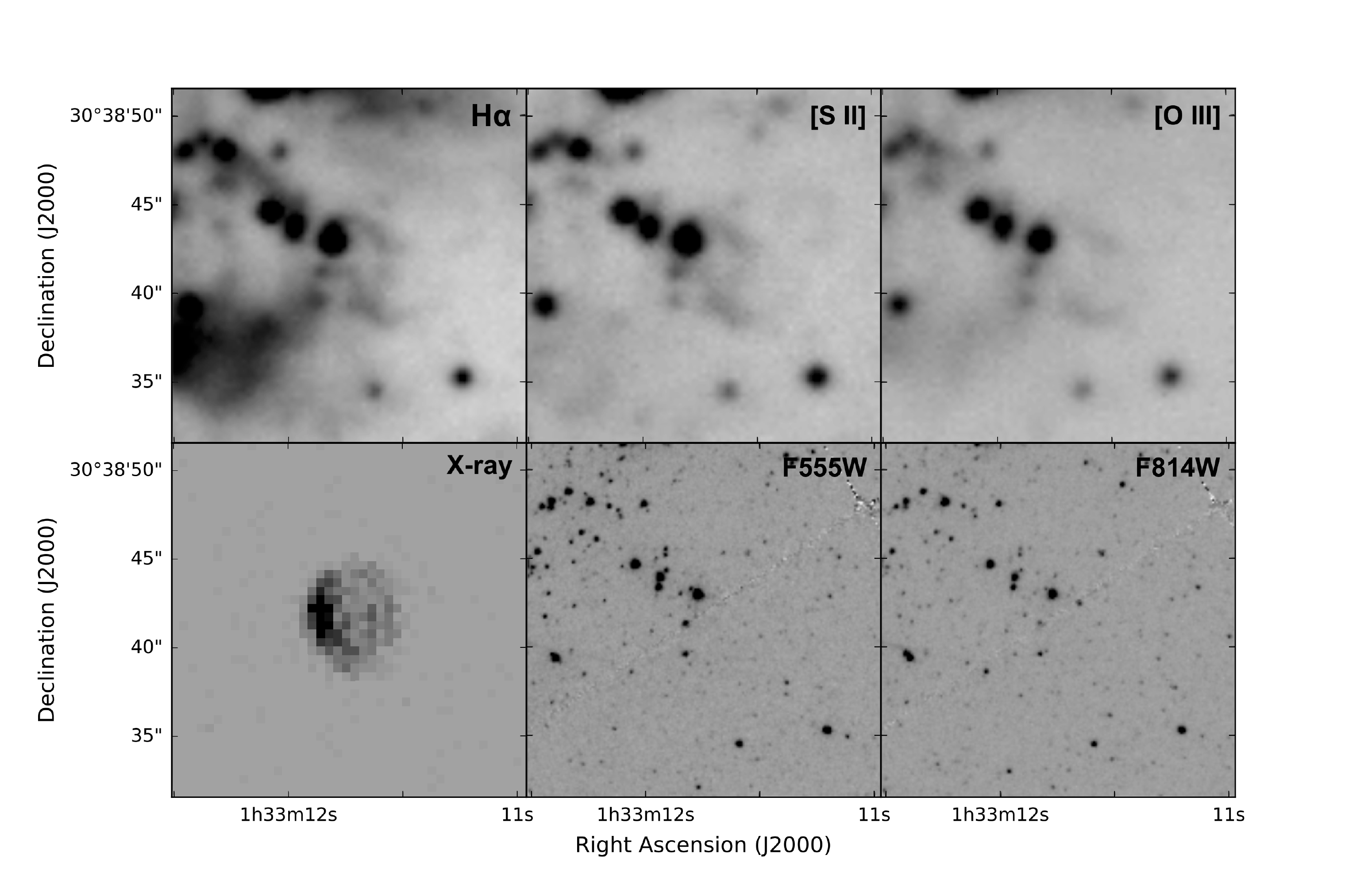}
    \caption{\textbf{No.102, 013311.75+303841.5 (L10-025).}  The \emph{HST} images
    were taken with the F555W ($V$) and F814W ($I$) filters in Program 11079 (PI: Bianchi).} 
    \label{fig:102SNR}
\end{figure}

\subsection{No.102, 013311.75+303841.5 (L10-025)(Figure \ref{fig:102SNR})}
With an X-ray luminosity over $10^{37}$ ergs s$^{-1}$, this SNR is one of the 
two brightest X-ray-emitting SNRs in M33 \citep{Tullmann2011}, and has thus been 
identified and well studied over the past two decades 
\citep{Long1996,Gordon1998,Gaetz2007,Long2010,Lee2014}. 

As analyzed and reported in detail by \citet{Gaetz2007}, this SNR is projected 
against the \ion{H}{2} region complex NGC\,592, and its shell structure is well 
resolved by $Chandra$ X-ray images. Figure \ref{fig:102SNR} shows that the 
SNR shell is detected in H$\alpha$, [\ion{S}{2}], and [\ion{O}{3}] images as well, 
although the optical shell morphology does not follow the X-ray shell closely.  
The SNR is located on the western periphery of an H$\alpha$ arc and the bright 
eastern rim of the X-ray shell abuts a dark cloud encompassed by the H$\alpha$ arc.  
It is possible that the bright X-ray emission is caused by the SNR shock running 
into the dense interstellar medium (ISM) near the molecular cloud (associated 
with the dark cloud).
Figure \ref{fig:102SNR} also shows three bright stellar sources to the northeast 
of the SNR, and the \emph{HST} F555W and F814W images clearly resolve the 
middle source into two stars.  The photometry of these stars reported by 
\citet{Massey2016} suggests that they are O supergiants.  The stellar and
 interstellar environments of this SNR are closely associated with massive stars, 
 suggesting a CC SN origin.

The most direct evidence of a SN's origin is provided by the SN ejecta abundance 
determined from X-ray spectra \citep{Hughes1995}.  The \emph{Chandra} X-ray 
spectra reported by \citet{Gaetz2007} show significant oxygen emission in the 
0.5--0.7 keV range relative to the iron emission peak. As the X-ray emission is 
distributed along the SNR shell rim, it most likely originates from the shock-heated 
ISM, rather than SN ejecta.  Thus, the X-ray spectra cannot constrain the SN 
progenitor of this SNR.

In summary, we suggest that this SNR has a CC SN origin.  While we cannot 
definitively exclude a Type Ia origin, the detection of [\ion{S}{2}] and [\ion{O}{3}] 
emission from the SNR clearly rejects it as a Balmer-dominated young Type Ia 
SNR nature.    

\begin{figure}[h]
    \centering
    \includegraphics[width=0.45\textwidth]{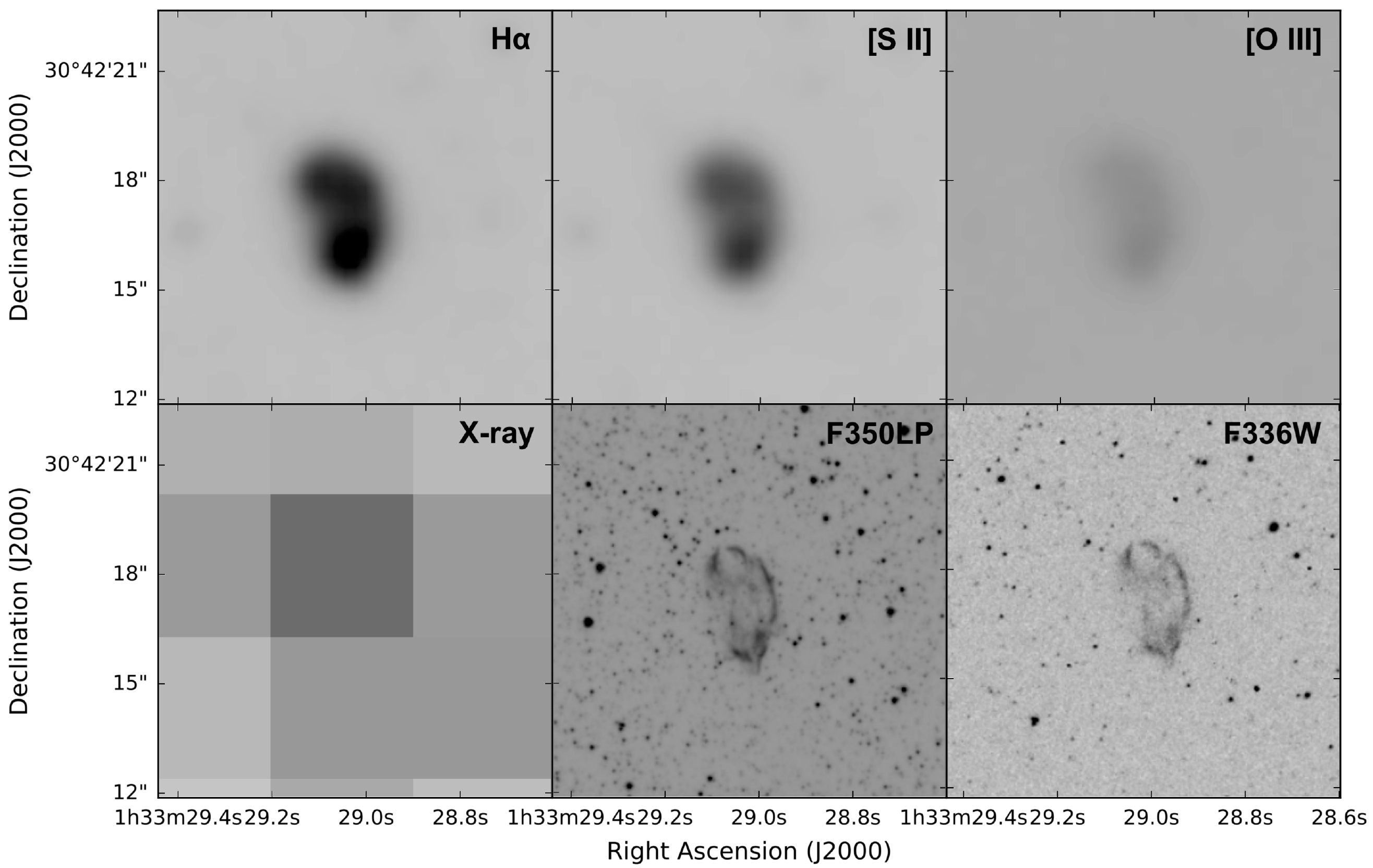}
    \caption{\textbf{No.184, 013329.04+304216.9 (L10-036, LL14-061).} 
    The \emph{HST} images were taken with the F350LP filter in 
     Program 13767 (PI: Trenti) and the F336W ($U$) filter in Program 14610
     (PI: Dalcanton).}
    \label{fig:184SNR}
\end{figure}

\subsection{No.184, 013329.04+304216.9 (L10-036, LL14-061)(Figure \ref{fig:184SNR})}
\begin{figure}[htbp]
    \centering
    \includegraphics[width=0.45\textwidth]{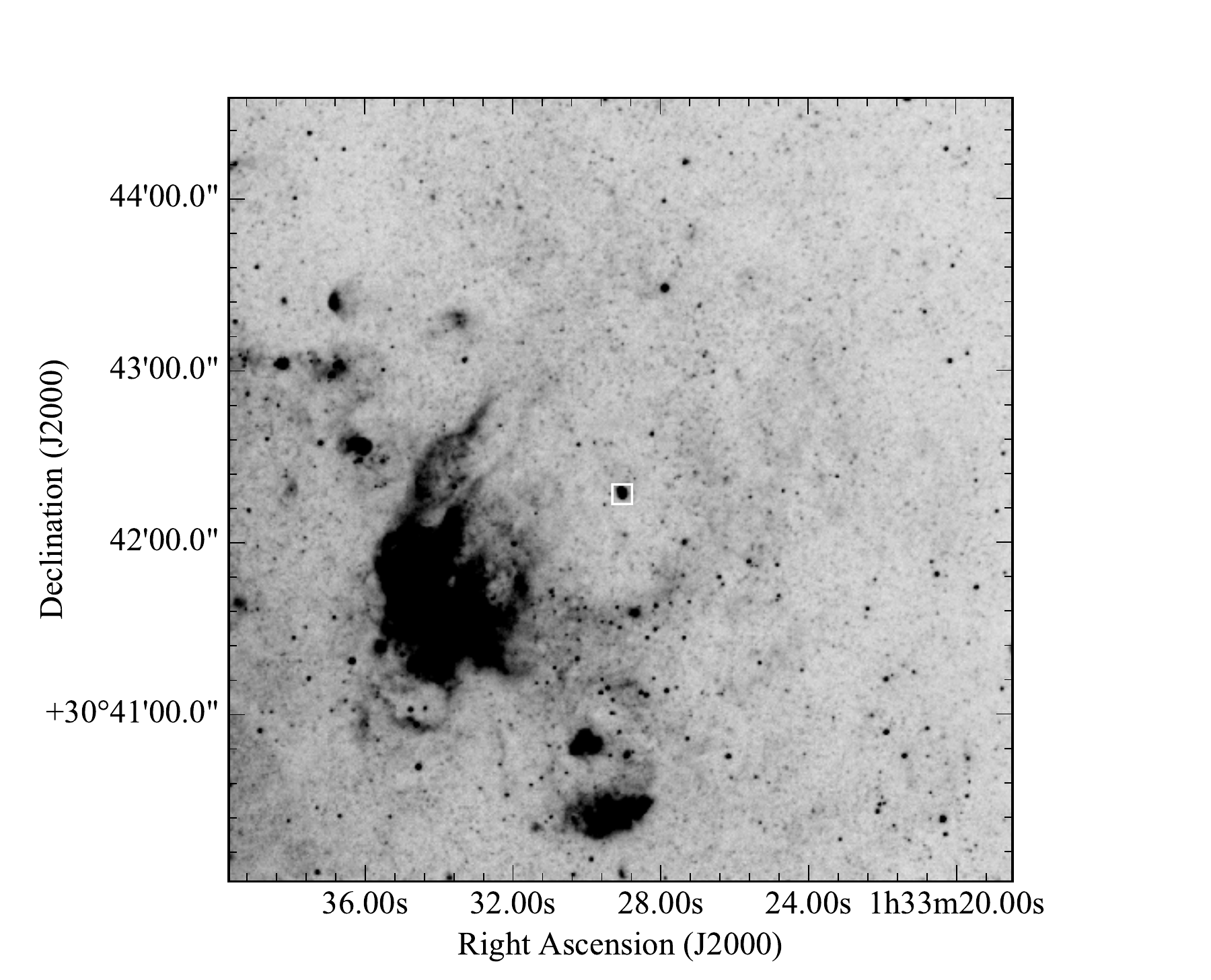}
    \caption{\textbf{No.184, 013329.04+304216.9 (L10-036, LL14-061)}.  
    The H$\alpha$ image shows that this SNR is projected within a superbubble 
    greater than 200 pc in size in NGC\,595. The white box marks the location 
    of this SNR.}
    \label{fig:184bubble}
\end{figure}

This SNR was first identified from H$\alpha$ and [\ion{S}{2}] images by
\citet{Sabbadin1979}. Its elongated shape, 11 pc along east-west and 18 pc 
along north-south, is clearly visible in the H$\alpha$, [\ion{S}{2}], and [\ion{O}{3}] 
images (Figure \ref{fig:184SNR}). The \emph{HST} F350LP and F336W ($U$) images 
have finally resolved the SNR shell, showing an opening to the east.
In a large-scale environment, as shown in Figure \ref{fig:184bubble}, it can be 
seen that this SNR is projected within a large superbubble more than 200 pc in 
size. The active star-forming region is located at the southeast corner of the
superbubble.  The apparent association with a star-forming region may suggest
that this SNR originates from a CC SN.  However, the SNR's shell morphology 
and large-scale environment are reminiscent of the Type Ia SNR N103B in the LMC 
\citep{Li2017}.  We do not know whether this SNR is physically located within the 
superbubble or merely projected against it, as N103B viewed along the direction 
toward the superbubble around NGC\,1850.  Furthermore, the F350LP and F336W 
filters contain Balmer lines as well as forbidden lines, and it is not clear whether the
Balmer lines and forbidden lines have different spatial distribution.  
As shown in Figure \ref{fig:LMC3SNRs}, 
some Balmer-dominated SNRs in the LMC do show knots of forbidden line emission 
with different spatial distribution from Balmer lines (Li et al.\ 2020, in preparation).
See N103B in Figure \ref{fig:LMC3SNRs} for an example and note the similarity in 
shell shape between SNR N103B and this SNR associated with Source 184.
In summary, the apparent interstellar environment of this SNR suggests that it is 
likely of CC SN origin; however, high-resolution \emph{HST} line images are 
needed to reveal the spatial distributions of Balmer and forbidden lines to conclude
definitively whether there is a Balmer-dominated shell in this SNR. 


\begin{figure}[h]
    \centering
    \includegraphics[width=0.45\textwidth]{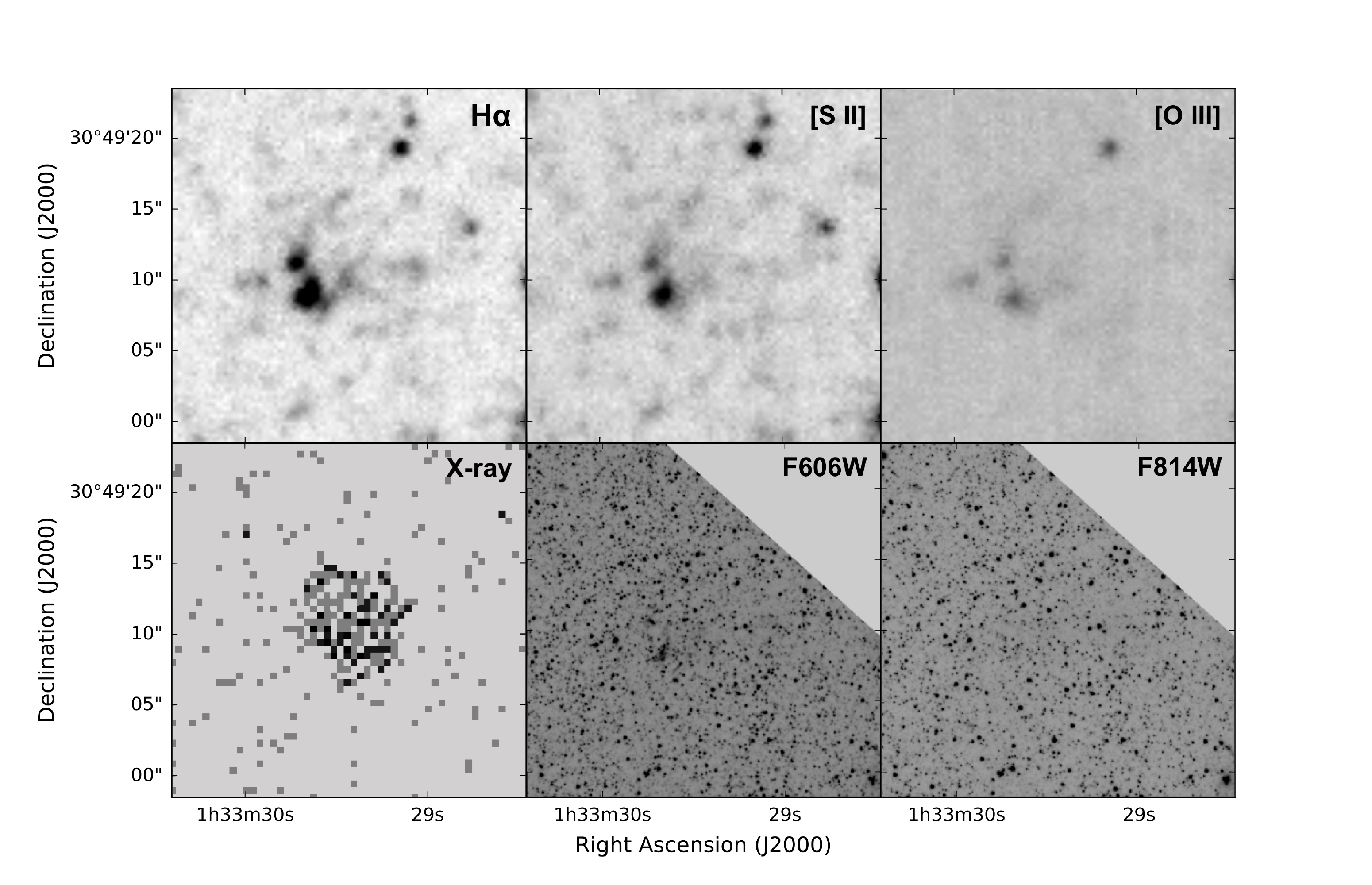}
    \caption{\textbf{No.188, 013329.45+304910.7 (L10-037, LL14-062).} 
    The \emph{HST} images were taken with the F606W ($V$) and F814W ($I$)
    filters in Program 9873 (PI: Sarajedini).}
    \label{fig:188SNR}
\end{figure}

\subsection{No.188, 013329.45+304910.7 (L10-037, LL14-062)(Figure \ref{fig:188SNR})}
The X-ray image of this SNR shows faint diffuse emission within a circular area 
with diameter of approximately 34 pc.  The H$\alpha$, [\ion{S}{2}], and [\ion{O}{3}] images 
show two bright patches of emission on the eastern side of the SNR.
The \emph{HST} F606W and F814W images detect diffuse emission from these two 
patches and reveal a cluster in the southern of the two.  The irregular optical morphology 
of this SNR and its possible association with a cluster with some ionizing power suggest 
that this SNR is of CC origin.  Furthermore, the presence of [\ion{S}{2}] and [\ion{O}{3}]
emission definitely rules out a Balmer-dominated Type Ia nature.

\begin{figure}[h]
    \centering
    \includegraphics[width=0.45\textwidth]{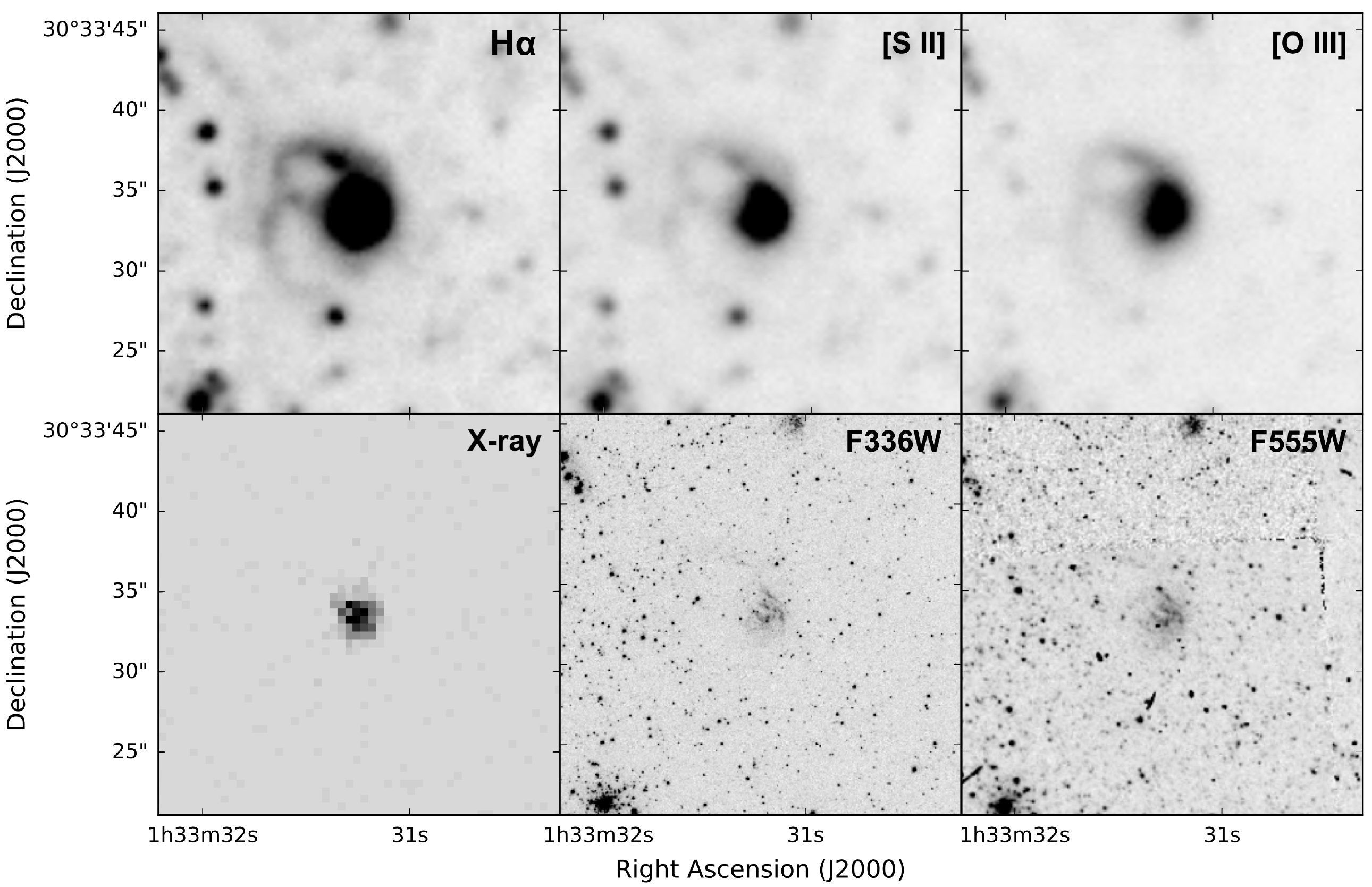}
    \caption{\textbf{No197, 013331.25+303333.4 (L10-039, LL14-067).} 
    The \emph{HST} images were taken with the F336W ($U$) in 
    Program 14610 (PI: Dalcanton) the F555W ($V$) filter in 
    Program 6038 (PI: Bianchi).}
    \label{fig:197SNR}
\end{figure}

\subsection{No.197, 013331.25+303333.4 (L10-039, LL14-067)(Figure \ref{fig:197SNR})}
Optical counterparts of X-ray source 197 are detected in H$\alpha$, [\ion{S}{2}], and [\ion{O}{3}] images.
A bright patch of emission is detected at the position of the X-ray source, and two fainter H$\alpha$ 
loops extend from the SNR location to the east.  The \emph{HST} F555W image has resolved the bright
patch of emission in the SNR into four roughly parallel filaments.  Furthermore, the \emph{HST} images 
resolve the object projected at $\sim$60 pc southeast from the SNR into a cluster.  
Lacking a well-defined shell structure is a characteristic often seen in CC SNRs, but not in
young Type Ia SNRs in the LMC; therefore, we suggest that the SNR associated with X-ray source 197 
is a CC SNR.

\begin{figure}[h]
    \centering
    \includegraphics[width=0.45\textwidth]{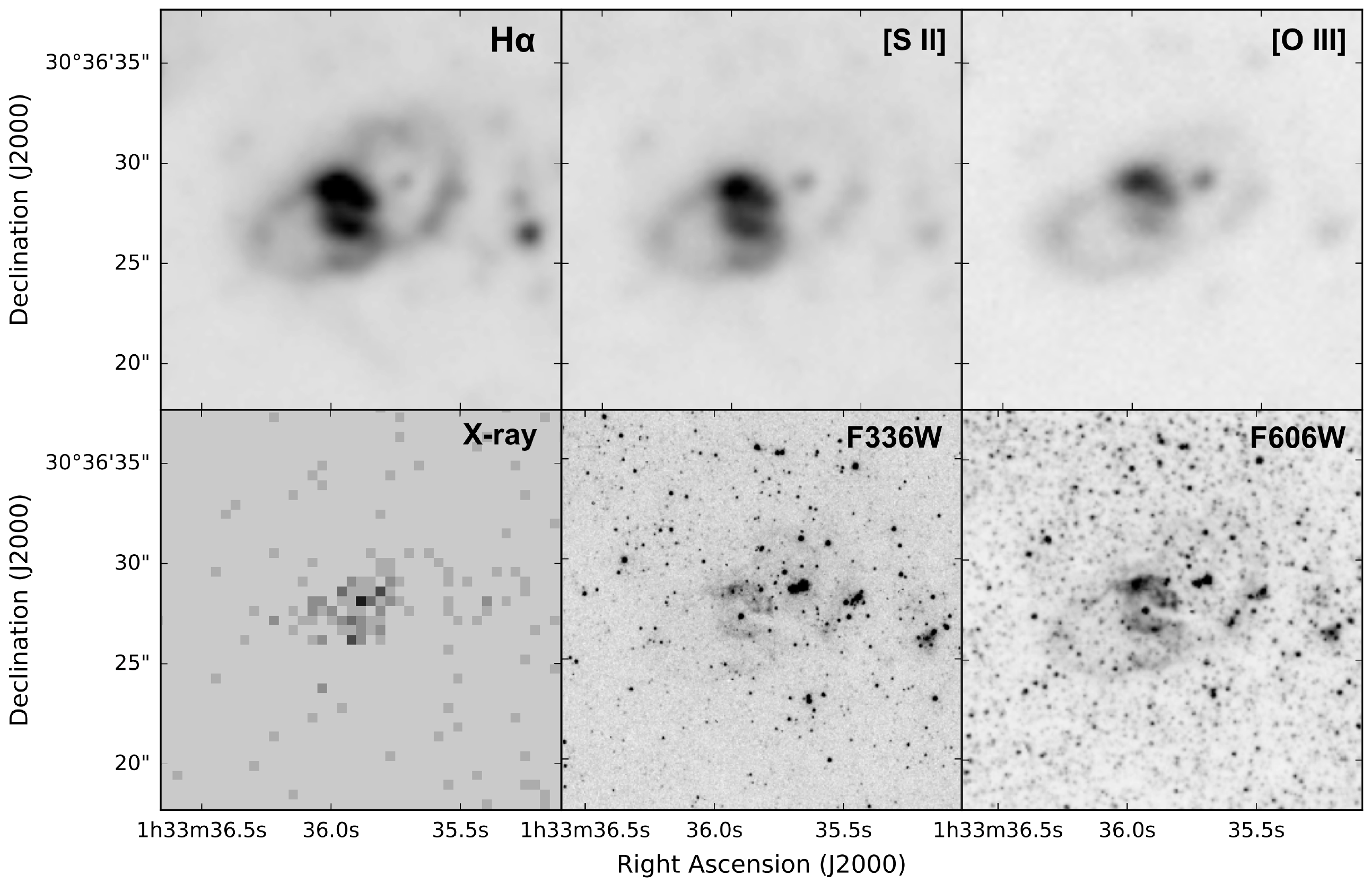}
    \caption{\textbf{No.236, 013335.90+303627.4 (L10-045, LL14-074).} 
    The \emph{HST} images were taken with the F336W ($U$) filter in Program 14610 
    (PI: Dalcanton) and the F606W ($V$) filter in Program 8090 (PI: Casertano).}
    \label{fig:236SNR}
\end{figure}

\subsection{No.236, 013335.90+303627.4 (L10-045, LL14-074)(Figure \ref{fig:236SNR})}

The H$\alpha$, [\ion{S}{2}] and [\ion{O}{3}] images all show bright emission at the location 
of the X-ray source, and from this bright central region two emission loops extending to 
the northwest and southeast directions and fainter arcs extending to the northeast and 
southwest.  The bright central emission is bisected by a narrow lane of lower surface 
brightness, and the overall morphology of the central bright emission and the two
brightest loops resembles two connected chain links.   The \emph{HST} F336W and 
F606W images show nicely that the narrow region of low surface brightness region in 
the bright emission region is a dust lane.  There is a star within the dust lane near the 
center of the bright central emission, but its photometric measurements from 
\citet{Massey2006, Massey2007, Massey2016} may be too contaminated by the 
nebular emission to be useful for spectral type estimates.
The association with the dust lane and the extended ionized loop structure indicates 
that the SNR is most likely a CC SNR.  Its [\ion{S}{2}] and [\ion{O}{3}] emission also 
rule out its being a Balmer-dominated young Type Ia SNR.

\begin{figure}[h]
    \centering
    \includegraphics[width=0.45\textwidth]{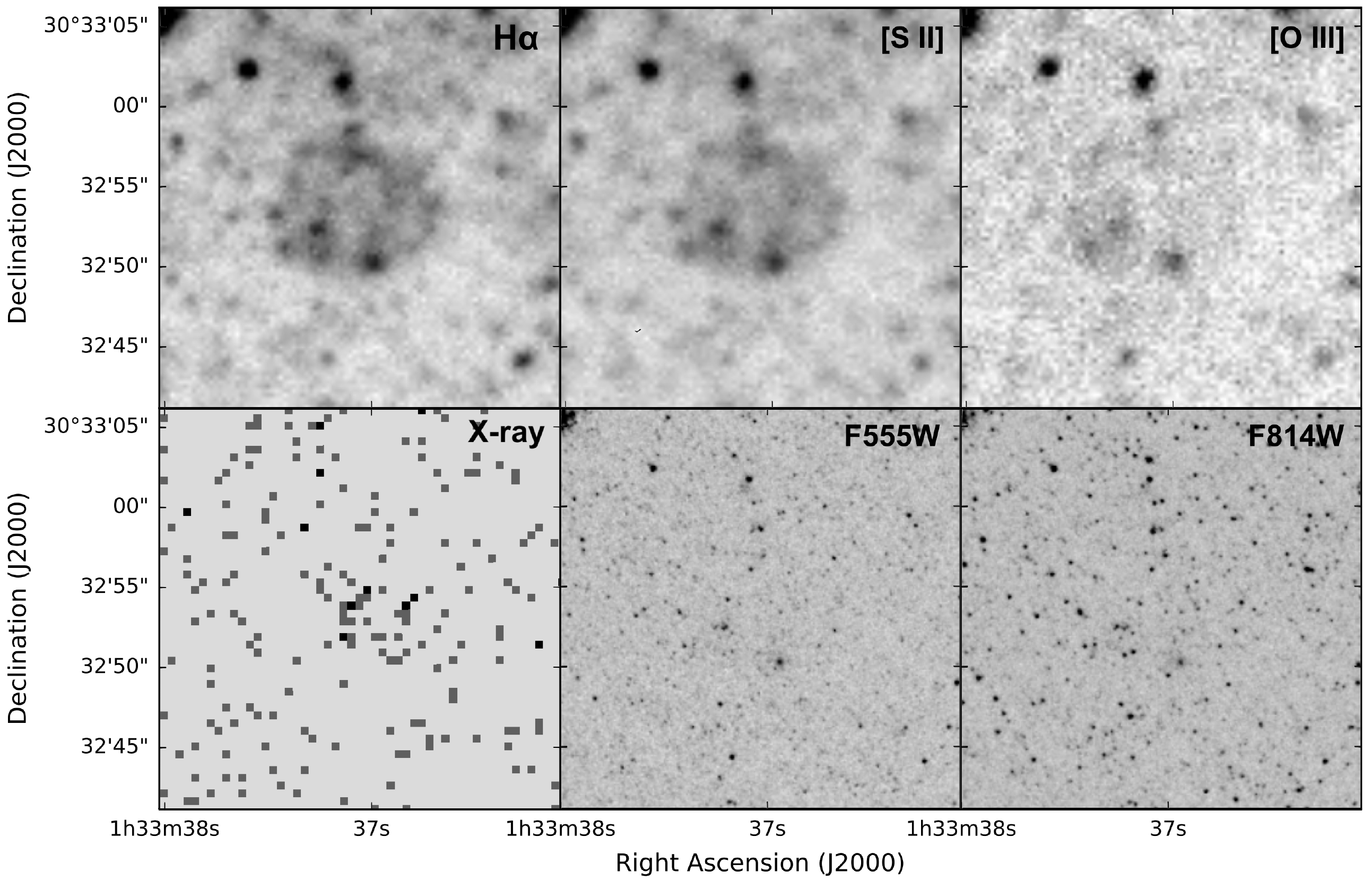}
    \caption{\textbf{No.245, 013337.08+303253.5 (L10-046, LL14-076).} 
    The \emph{HST} images were taken with the F555W ($V$) and F814W ($I$) filters
    in Program 11079 (PI: Bianchi).}
    \label{fig:245SNR}
\end{figure}

\subsection{No.245, 013337.08+303253.5 (L10-046, LL14-076)(Figure \ref{fig:245SNR})}
This SNR is located along the prominent southern spiral arm of M33.  The nearest large 
star-forming region is located at about 90 pc in the southeast.
The SNR has a weak limb-brightening in the H$\alpha$ image, indicating a shell structure; 
however, there is still wide-spread diffuse emission within the shell boundary.  
The [\ion{S}{2}] image shows a SNR morphology similar to that 
in H$\alpha$.  The [\ion{O}{3}] image shows slightly different surface brightness 
distribution and a less well-defined shell rim on the west side.  The X-ray emission is 
distributed near the center of the SNR shell, as opposed to along the shell rim.  
This SNR is almost the largest among the nine X-ray-bright SNRs discussed in this paper, 
but it has the highest unabsorbed X-ray luminosity.  This high unabsorbed X-ray luminosity 
probably results from a very high absorption correction, as this SNR has the lowest plasma 
temperature and the highest absorption column among the nine SNRs in Table \ref{tab:table2}.
The H$\alpha$ surface brightness of this SNR is so low that its \emph{HST} F555W and 
F814W images optimized to observe stars could not detect any line emission from this SNR.
We are not certain whether this is a CC or Type Ia SNR.  The presence of [\ion{O}{3}] and
[\ion{S}{2}] emission indicates that this SNR is definitely not a young Balmer-dominated Type Ia SNR.

\begin{figure}[h]
    \centering
    \includegraphics[width=0.45\textwidth]{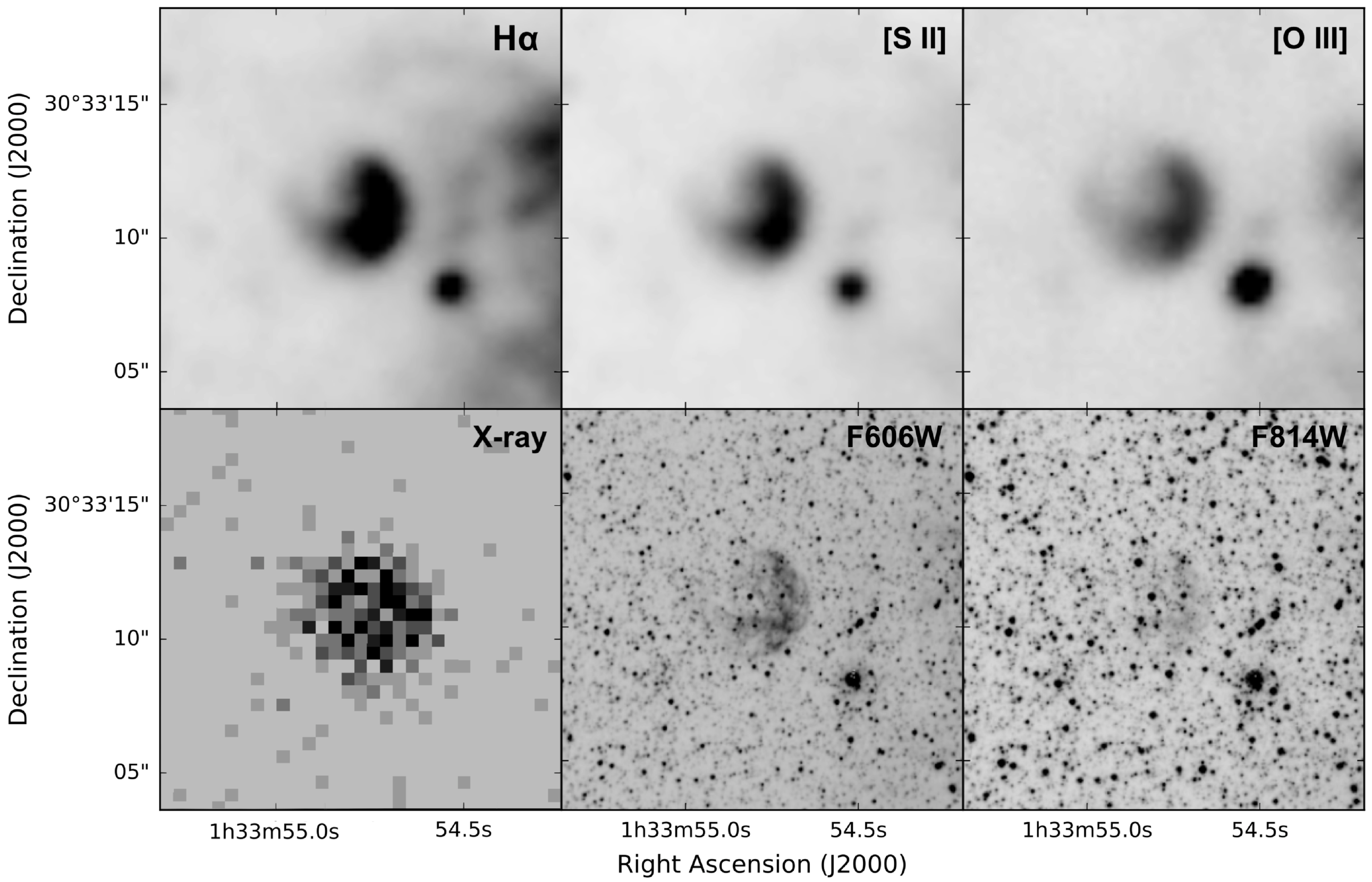}
    \caption{\textbf{No.334, 013354.91+303310.9 (L10-071, LL14-107).} 
    The \emph{HST} images were taken with the F606W ($V$) and F814W ($I$)
    filters in Program 10190 (PI: Garnett).}
    \label{fig:334SNR}
\end{figure}

\subsection{No.334, 013354.91+303310.9 (L10-071, LL14-107)(Figure \ref{fig:334SNR})}
The X-ray image of this SNR appears larger than the optical counterpart, but this is likely 
caused by the large point-spread-function of the SNR's off-axis position in the ACIS-I detector. 
The SNR is detected in H$\alpha$, [\ion{S}{2}] and [\ion{O}{3}] images.  
The \emph{HST} F606W image resolves the filamentary structure of the SNR that is detected 
only over the western hemisphere.  
The strongest nebular line in the F606W band is the H$\alpha$+[\ion{N}{2}] line. 
The image in the F606W band most likely represents an H$\alpha$ image.  
The nebular morphology is very similar to that of the CC SNR in N4, an \ion{H}{2} region 
ionized by an OB association in the LMC \citep{Chu1997}.  The SNR 
associated with X-ray source 334 is located only 40 pc east from a cluster that is embedded 
in diffuse interstellar H$\alpha$ emission (visible on the western edge of the
panels in Figure \ref{fig:334SNR}). 
The low surface brightness of the diffuse H$\alpha$ emission implies that the cluster is 
relatively evolved (10$^7$ yr or older) and has no powerful O stars left.  If the SN progenitor 
had been ejected by this cluster, it would probably have been a B star.  Based on this SNR's 
morphology and its interstellar and stellar environment, we suggest that it is a CC SNR.  

\begin{figure}[h]
    \centering
    \includegraphics[width=0.45\textwidth]{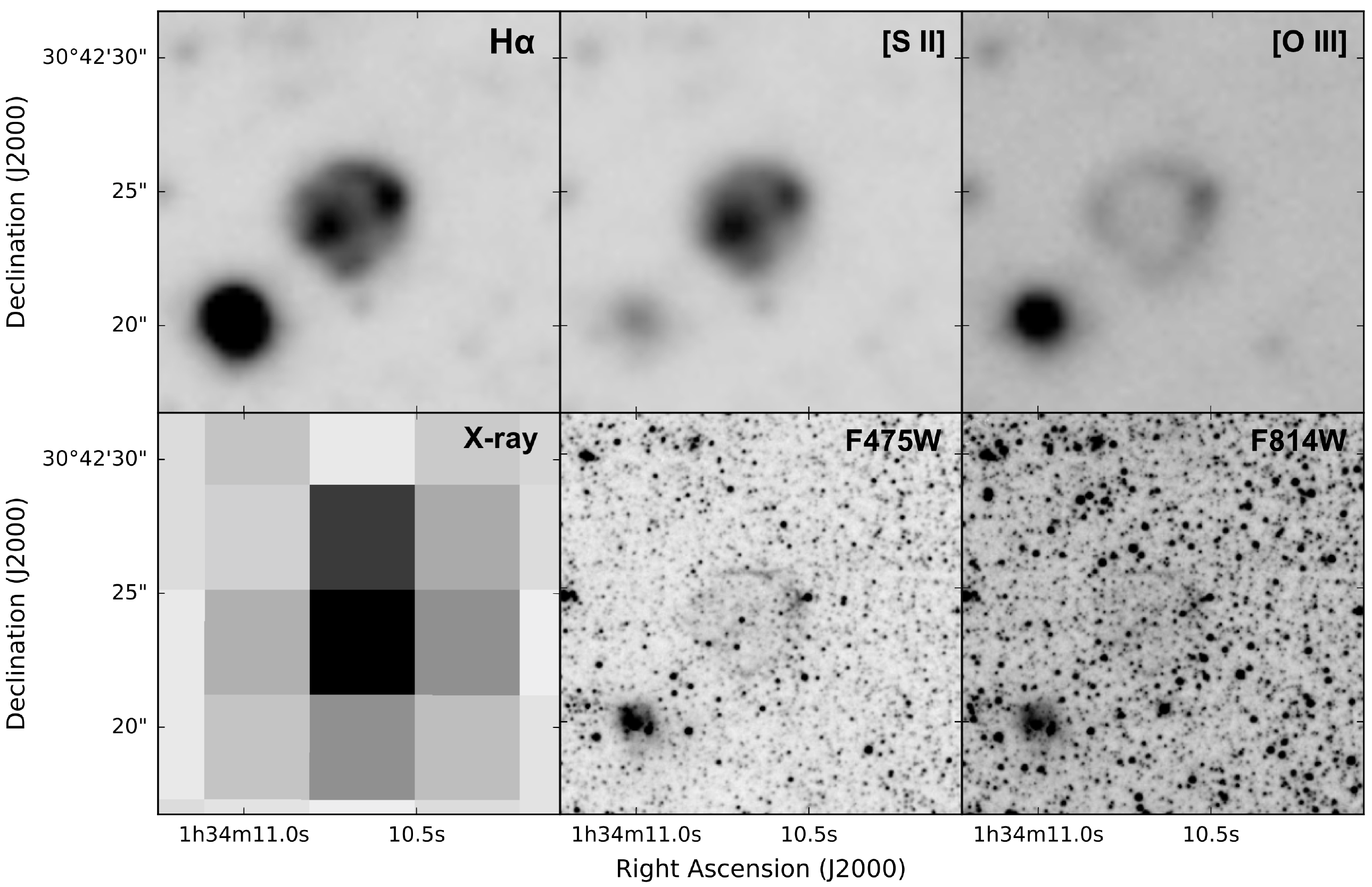}
    \caption{\textbf{No.427, 013410.69+304224.0 (L10-096, LL14-140).} 
    The \emph{HST} images were taken with the F475W ($B$) and F814W ($I$)
    filters in Program 14610 (PI: Dalcanton).}
    \label{fig:427SNR}
\end{figure}

\subsection{No.427, 013410.69+304224.0 (L10-096, LL14-140)(Figure \ref{fig:427SNR})}
This SNR shows a shell structure best in the [\ion{O}{3}] image, with a bright spot on the 
northwest rim (about 60$^\circ$ from the north).
The H$\alpha$ and [\ion{S}{2}] images show more diffuse emission in the central region, 
and an additional bright spot on the southeast rim, which is most likely caused by a red 
star that is best seen in the \emph{HST} F814W image.  The \emph{HST} F475W image
includes the H$\beta$ and [\ion{O}{3}] line emission and thus detects the SNR shell.  
This high-resolution image shows a sharp shell filament in the northern rim, but much 
more diffuse for the rest of the shell rim.  This shell morphology is very different from 
those of the Balmer-dominated Type Ia SNRs in the LMC.  Although there are no large 
\ion{H}{2} complexes in the vicinity of this SNR, a bright \ion{H}{2} region is projected at 
$\sim$23 pc away.  We suggest that this is likely a CC SNR, and definitely
not a Balmer-dominated Type Ia SNR.

\section{Discussion}
Using the young Balmer-dominated Type Ia SNRs in the LMC as a template, we have 
searched M33 for similar objects.  As the Type Ia SN rate is proportional to the stellar mass 
of a galaxy \citep{Sullivan2006} and M33 has 2--3 times the stellar mass of the LMC 
(see Table \ref{tab:table3} and references therein), we expect at least a few 
Balmer-dominated Type Ia SNRs in M33; however, we did not find any unambiguous 
candidate!  This result is very puzzling.  To solve this puzzle, we consider below effects 
of observational bias, density and ionization fraction of the ISM, Type Ia
SN rate estimated from star formation history and delay time distribution (DTD),
and metallicity.

\begin{deluxetable*}{lcccll}
\tablecaption{Comparison between the LMC and M33 \label{tab:table3}}
\tablehead{
\colhead{}                      & \colhead{Units} & \colhead{LMC} & \colhead{M33} & \colhead{References of LMC} & \colhead{References of M33}}
\startdata
Total Baryonic Mass      & $M_{\odot}$ &  3.2$\times$10$^{9}$                  & 6.2--9.2$\times$10$^{9}$ & \cite{Marel2002} & \cite{Corbelli2003}\\
HI mass                         & $M_{\odot}$ & (4.8$\pm$0.2)$\times$10$^{8}$  & 1.4$\times$10$^{9}$       & \cite{Staveley2003} & \cite{Gratier2010}\\
H$_2$ mass (CO line)  & $M_{\odot}$  & $\sim$ 5$\times$10$^{7}$           & 3.3$\times$10$^{8}$       & \cite{Fukui2008} & \cite{Gratier2010}\\
Total gas mass             & $M_{\odot}$   & $\sim$ 5.3$\times$10$^{8}$        & 1.73$\times$10$^{9}$     &   ...                     & ... \\
Recent SFR                 & $M_{\odot}$ yr$^{-1}$ &  0.21                               & 0.25                                 & \cite{Chomiuk2011} & \cite{Chomiuk2011}\\
Total SNRs                   &                        & 59                                               & 199                                  & \cite{Maggi2016} & \cite{Lee2014}\\
Confirmed Balmer-dominated SNRs &   & 5                                                 & 0                                     & \cite{Ou2018}     & this paper
\enddata
\end{deluxetable*}

\subsection{Observational Bias Effect}
\begin{figure}[h]
    \centering
    \includegraphics[width=0.4\textwidth]{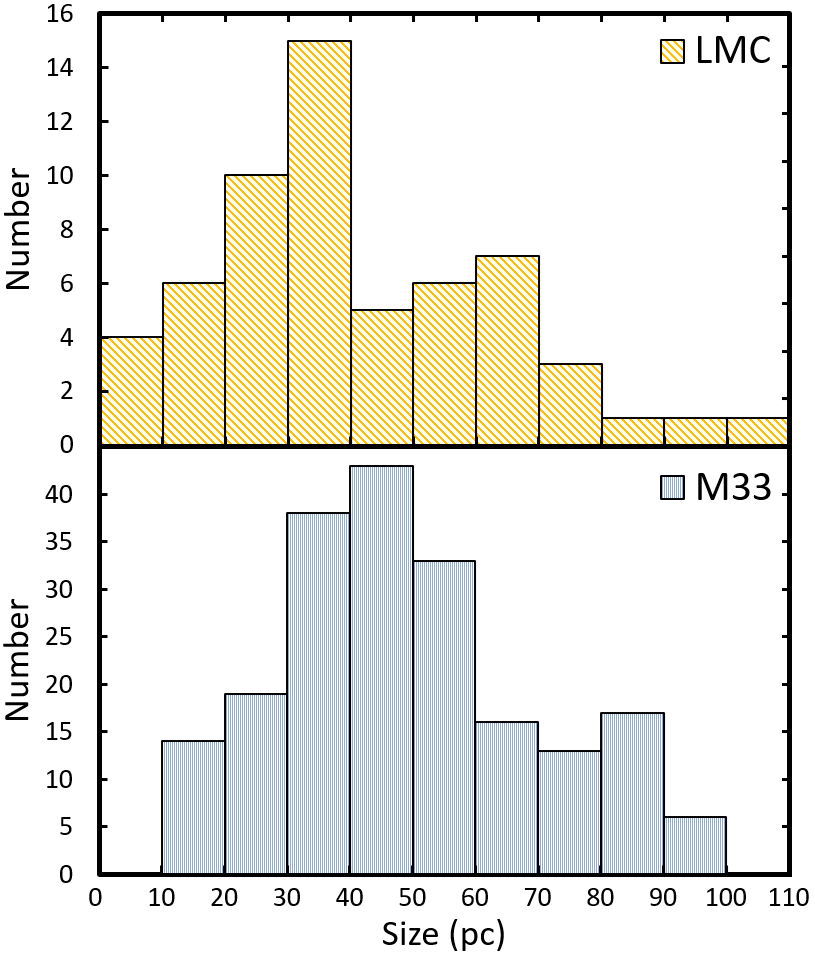}
    \caption{Size distributions of known SNRs in the LMC and M33. The optical sizes of the 
    59 known SNRs in the LMC are from \citet{Ou2018}, and those of 199 known SNRs in 
    M33 are from \citet{Lee2014}.}
    \label{fig:SizeDistribution}
\end{figure}

Surveys of SNRs in the LMC have confirmed at least $\sim$60 objects \citep{Maggi2016, Bozzetto2017} 
and among them five are Balmer-dominated Type Ia SNRs, while surveys of SNRs in M33 have confirmed 
$\sim$200 SNRs \citep{Lee2014} and zero Balmer-dominated Type Ia SNRs (this paper).   
{The M33-to-LMC SNR number ratio, 200/60 $\sim$ 3.3, is not very different from their galactic stellar 
mass ratio, $\sim$ 2--3.  The size distributions of known SNRs in the LMC and M33 are shown in 
Figure \ref{fig:SizeDistribution}.  The deficiency of small SNRs in M33, especially those smaller than 
10 pc in size, is a clear indication of incompleteness of SNR surveys for M33 due to the limiting angular 
resolution of ground-based optical observations, as 10 pc corresponds to an angular size of 2\farcs5.}  

Our new search for young Balmer-dominated Type Ia SNRs in M33 is not subject to the size versus 
angular resolution effect because we use X-ray luminosity to make the initial selection of candidates. 
{There is nevertheless a real difference in the X-ray luminosity function between M33 and the LMC.
The cumulated X-ray luminosity function of M33 SNRs has been compared with that of the LMC SNRs
using \emph{XMM-Newton} and \emph{Chandra} observations \citep{Maggi2016, Long2010, Garofali2017}.
It is evident that the LMC has more X-ray-luminous SNRs than M33; for example, the LMC has 14 SNRs
more luminous than 5$\times$10$^{35}$ ergs~s$^{-1}$, while M33 has only nine.  This difference is 
real and not caused by observational bias effects.  
It is interesting to note that among these 14 luminous LMC SNRs five are Balmer-dominated Type Ia SNRs,
and among the 9 luminous M33 SNRs none is Balmer-dominated.  
These numbers appear to suggest that the difference between the LMC and M33 in the 
cumulated X-ray luminosity function is caused by an absence of luminous Balmer-dominated 
Type Ia SNRs in M33.}

\subsection{Interstellar Density Effect}

M33 has a total gas mass of $\sim1.73\times10^9$ $M_\odot$ and a radius of 8.2 kpc \citep{Gratier2010}, 
while the LMC has a total gas mass of $\sim5.3\times10^8$ $M_\odot$ and a radius of 5.2 kpc 
\citep{Staveley2003,Fukui2008}.  Assuming a similar gas scale height for M33 and the LMC, the average 
gas density in M33 is about 20\% higher than that in the LMC.  This difference in average interstellar gas 
densities means that X-ray emission from shocks into the ISM would be higher for Type Ia SNRs in M33 
than those in the LMC.   Thus, the density of the ISM cannot be responsible for the missing X-ray-bright 
Balmer-dominated Type Ia SNRs in M33.

\subsection{Ionization Fraction of the ISM Effect}

If a Type Ia SN explodes in an ionized ISM, it may not have a Balmer-dominated spectrum. 
The total ionizing flux available in a galaxy is proportional to the current star formation rate, which 
is {$0.25\pm0.1$ $M_\odot$ yr$^{-1}$ for M33 and $\sim$0.21 $M_\odot$  yr$^{-1}$ for 
\citep{Chomiuk2011}. } The total gas mass of M33 is $\sim$3 times the total gas mass of the LMC.  
Therefore, with a high total gas mass but a comparable ionizing flux, M33 would have a larger 
fraction of neutral ISM, and we do not expect it to prohibit the formation of Balmer-dominated 
Type Ia SNRs in M33.

\subsection{Type Ia SN Rate Expected from Star Formation History}

\begin{figure}[h]
    \centering
    \includegraphics[width=0.4\textwidth]{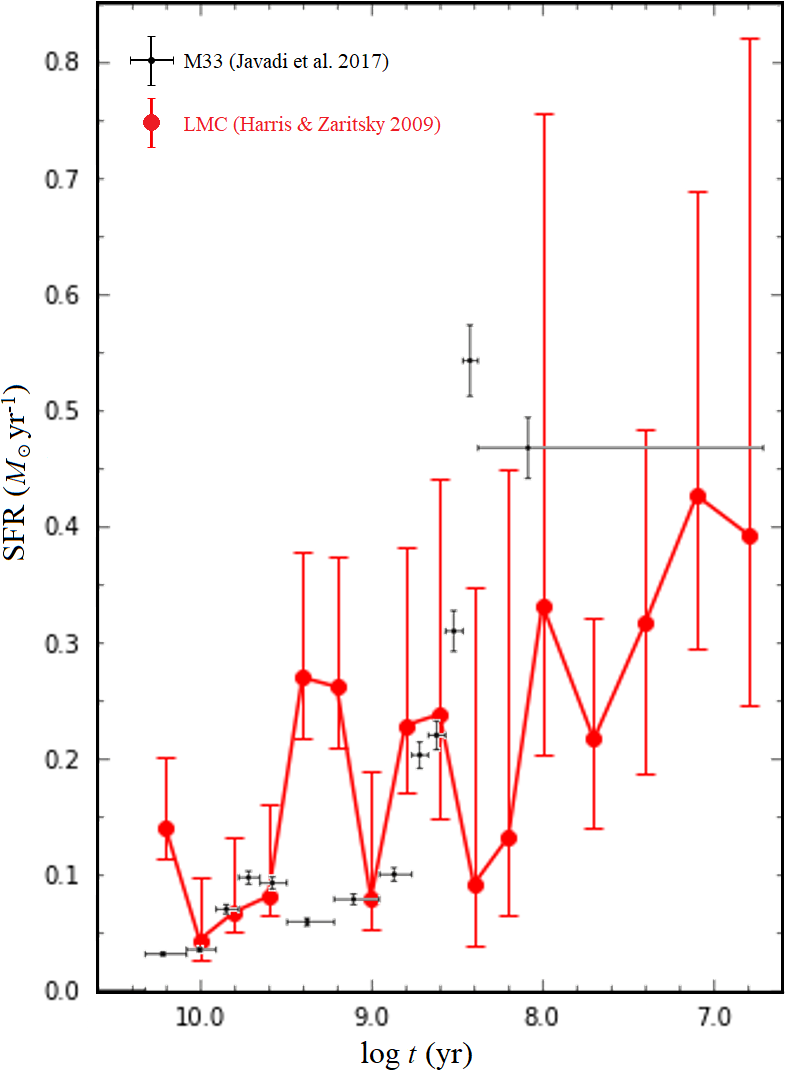}
    \caption{Star formation history of the LMC and M33. The SFH of the LMC is from \cite{Harris2009} 
    and the SFH of M33 is from \cite{Javadi2017}.}
    \label{fig:SFH}
\end{figure}

\begin{figure}[h]
    \centering
    \includegraphics[width=0.4\textwidth]{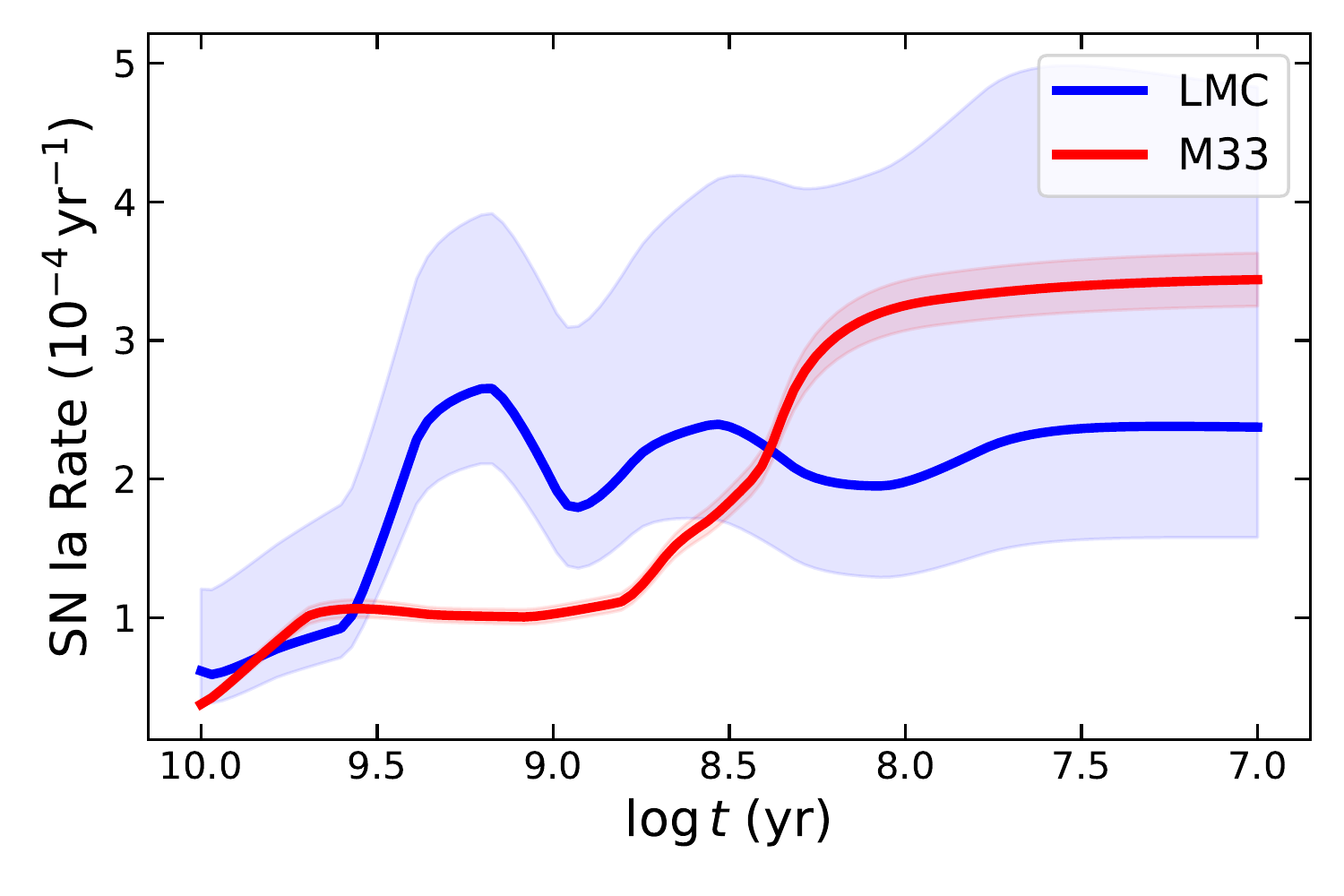}
    \caption{Expected Type Ia SN rates for the LMC and M33. }
    \label{fig:SNIa_rate}
\end{figure}

{For a single burst of star formation, the rate of Type Ia SNe is a function of ``delay time",
the time lapse from the star formation to SN explosion.  This function, called delay-time
distribution (DTD) function, is expressed in a power law, $\propto t^\alpha$,
where $t$ is the delay time and greater than 40 Myr, and $\alpha$ has been 
determined from observations.  (The DTD function is zero at $t <$ 40 Myr.)
For a galaxy with varying star formation rate in the past, the expected Type Ia SN 
rate can be derived by convolving the star formation history (SFH) with the DTD 
function \citep{Maoz2012}.  The SFH of the LMC has been determined by 
\citet{Harris2009} and the SFH of M33 has been determined by \citep{Javadi2017}, 
as shown in Figure~\ref{fig:SFH}, where star formation rates (SFR) are plotted 
as a function of look-back time.  We have interpolated each SFH to obtain a 
continuous function of $t$ for SFR and convolved it with a DTD function using 
$\alpha \sim -1.07$ and the Hubble-time-integrated SN Ia production 
efficiency of 1.3$\pm$0.1 Type Ia SN per 1000 M$_{\odot}$ of formed 
stellar mass determined by \citet{Maoz2017}.
The resulting Type Ia SN rates of the LMC and M33 are plotted in Figure~\ref{fig:SNIa_rate}.
Given the SFH of the LMC and M33, we expect M33 to have 50\% more Type Ia SNRs than
the LMC; however, we did not detect any X-ray-bright Type Ia SNRs in M33.}

\subsection{Metallicity Effect}

The metallicity of a galaxy affects Balmer-dominated Type Ia SNRs in two respects.  First, 
the Type Ia SN rate of a galaxy depends on metallicity, $Z$ \citep{Kistler2013}. The rate is 
proportional to $Z^{-0.3}$ to $Z^{-0.5}$ for single-degenerate SNe and $Z^{-0.4}$ for 
double-degenerate SNe.  The stellar abundances of M33 have a bimodal distribution, 
with [Fe/H] peaking near $-$1.3 and $-$1.7, while the distribution of LMC stellar 
abundances peaks at [Fe/H] $\sim -1.15$ \citep{Cioni2009}.  We can adopt an approximate 
metallicity dependence of $Z^{-0.4}$ for the Type Ia SN rate.  For the stellar masses and 
metallicities of M33 and the LMC, the Type Ia SN rate of M33 should be at least $\sim$3 times 
as high as the LMC.  This metallicity effect does not solve the puzzle of not detecting X-ray-bright
Balmer-dominated Type Ia SNRs in M33.

The second metallicity effect is on the X-ray emissivity that is directly proportional to the 
metal content of the plasma. X-ray emission from a young Type Ia SNR can originate from the 
forward shocks into ISM and the reverse shocks into the SN ejecta.  The abundance of the 
SN ejecta is determined by the nucleosynthesis in the SN explosion and depends on the SN 
explosion mechanism.  Without a priori knowledge of the differences in Type Ia SN populations 
between M33 and the LMC, we assume the Type Ia SN ejecta abundances are similar between 
M33 and the LMC.  The X-ray emission associated with the forward shocks is from shocked ISM.  
The oxygen abundance 12+log(O/H) of M33 is in the range of 8.1 to 8.5 \citep{Rosolowsky2008}, 
while the LMC has $\sim$8.35 \citep{Russell1992}.  Within a factor of 2, the interstellar abundances 
of M33 and the LMC are very similar.  There is no indication that the metallicities of M33 and the 
LMC are different enough to cause large differences in the X-ray emissivity of Type Ia SNRs.

\section{Summary and Conclusion}
{ Conventional surveys of SNRs using [\ion{S}{2}]/H$\alpha$ ratios as diagnostics would miss 
the young Balmer-dominated Type Ia SNRs.  We have thus applied a new methodology to search for 
Balmer-dominated Type Ia SNRs in M33: using $L_{\rm X} \ge 5\times10^{35}$ ergs s$^{-1}$ 
to select thermal X-ray sources, examining their optical counterparts in H$\alpha$, and checking 
whether they are detected in [\ion{S}{2}] and [\ion{O}{3}] images.  While there are nine known
SNRs in M33 meet the X-ray luminosity criterion, none of them are Balmer-dominated.

The LMC hosts five Balmer-dominated Type Ia SNRs, and all of them have X-ray luminosities 
greater than 8$\times$10$^{35}$ ergs s$^{-1}$ and thermal X-ray emission.  Among the five
smallest known Type Ia SNRs in the Galaxy, the Tycho and Kepler SNRs have X-ray properties 
similar to those in the LMC, while G1.9+0.3, SN1006, and RCW86 have X-ray luminosities lower
than 5$\times$10$^{35}$ ergs s$^{-1}$ and show nonthermal X-ray emission.  
Our X-ray luminosity selection criterion for Balmer-dominated SNRs in M33 aims at the 
X-ray-bright SNRs similar to those in the LMC or Kepler and Tycho in the Galaxy.

The absence of X-ray-bright Balmer-dominated Type Ia SNRs in M33 is surprising, because
M33's stellar mass is 2--3 times the LMC's mass and we expect at least a few X-ray-bright
Balmer-dominated Type Ia SNRs in M33.
We have computed the Type Ia SN rates expected from the SFH and SN Ia DTD 
function, and find that Type Ia SN rate of M33 should be 1.5 times that of the LMC,
contrary to what we have found. 
We have also considered observational biases, interstellar densities and ionization conditions, 
and metallicity effects, but none of these effects can explain the absence of X-ray-bright 
Balmer-dominated Type Ia SNRs in M33.   

It is intriguing that the Galaxy, LMC, and M33 exhibit different Type Ia SNR population --
The Galaxy has both X-ray-bright and X-ray-faint Type Ia SNRs, the LMC does not have the
X-ray-faint ones, while M33 does not have the X-ray-bright  ones.}

\acknowledgments 
We are very thankul to the anonymous reviewer who has made very useful comments and suggestions
to improve this paper.  This research is supported by grants MOST 107-2119-M-001-018 
and MOST 108-2112-M-001-045 from the Ministry of Science and Technology 
of Taiwan, Republic of China.



\end{document}